\documentclass[useAMS,usenatbib, 
]{mnras}
\usepackage{natbib}
\usepackage{graphicx}
\usepackage{epsfig}
\usepackage{amssymb}
\usepackage{amsmath}
\usepackage{natbib}
\usepackage{times}

\usepackage[english]{babel}

\title[Nature of the dust streaming instability]{On the nature of the resonant drag instability of dust 
streaming in protoplanetary disc}
\author[V.V. Zhuravlev]{V. V. Zhuravlev$^{1}$\thanks{E-mail:
zhuravlev@sai.msu.ru} \\
$^{1}$Sternberg Astronomical Institute, Lomonosov Moscow State University, Universitetskij pr., 13, Moscow 119234, Russia}

\begin{document}

\date{
}

\pagerange{\pageref{firstpage}--\pageref{lastpage}} \pubyear{2017}

\maketitle

\label{firstpage}

\defcitealias{squire_2018}{SH}

\begin{abstract}

The recently discovered resonant drag instability (RDI) of dust streaming in protoplanetary disc is considered 
as the mode coupling of subsonic gas-dust mixture perturbations. This mode coupling is coalescence of two modes 
with nearly equal phase velocities: inertial wave (IW) having positive energy and a streaming dust wave (SDW) 
having negative energy as measured in the frame of gas environment being at rest in vertical hydrostatic equilibrium. 
SDW is a trivial mode produced by the bulk streaming of dust, which transports perturbations of dust density. 
In this way, settling combined with radial drift of the dust makes possible coupling of SDW with IW and the onset of the instability. 
In accordance with the concept of the mode coupling, RDI growth rate is proportional to the square root of the 
coupling term of the dispersion equation, which itself is proportional to mass fraction of dust, $f\ll 1$. 
This clarifies why RDI growth rate $\propto f^{1/2}$. 
When SDW has positive energy, its resonance with IW provides an avoided crossing instead of the mode coupling.
In the high wavenumber limit RDI with unbounded growth rate $\propto f^{1/3}$ is explained by the triple mode coupling, 
which is coupling of SDW with two IW. 
It coexists with a new quasi-resonant instability accompanied by bonding of two oppositely propagating low-frequency IW.
The mode coupling does not exist for dust streaming only radially in a disc. 
In this case RDI is provided by the obscured mechanism associated with the inertia of solids.

\end{abstract}


\section{Introduction}

Relative motion of gas and dust in protoplanetary discs drives the generic dynamical instability of gas-dust mixture 
due to interaction between two components via aerodynamic drag. \citet{squire_2018_0} recognised that instability arises 
each time there is synchronization between the streaming dust motion and the wave of any nature propagating in gas environment.
They termed this type of instability as resonant drag instability (RDI). This gave a new look at the previously known 
streaming instability of \citet{youdin-goodman-2005}, which arises due to the radial drift of solids in the midplane of protoplanetary disc.
Currently, the streaming instability is a candidate for the solids growth mediator between pairwise sticking of particles 
and gravitational instability of dust-laden sub-disc via classic Safronov-Goldreich-Ward mechanism, see \citet{safronov-1972} and \citet{goldreich-ward-1973}, which ends with formation of planetesimals.
This view on the role of the streaming instability was established in the work by \citet{johansen-2007-apj-1} and \citet{johansen-2007-apj-2} who investigated its non-linear regime, and subsequently by \citet{johansen-2007-nat} who demonstrated that the resulting dust 
overdensities are gravitationally unstable.
However, the subsequent simulations have shown that this scenario is sensitive to dust fraction: it requires the initial metalicity of disc be a few times higher than the solar metalicity, otherwise turbulence caused by the streaming instability leads to stirring 
rather than clumping of dust, see e.g. \citet{johansen-2009}, \citet{bai-stone-2010}. This constraint becomes more severe due to 
limitations for grain size. 

Along with the known mechanisms of dust pile-up in long-living turbulent structures of disc such as zonal flows and anticyclonic vortices, 
for the review see e.g. \citet{chiang-youdin-2010}, \citet{johansen-2014}, \citet{johansen-2016}, the local amount of dust in a disc 
can potentially be enhanced through the new vigorous instability found by \citet{squire_2018} (below \citetalias{squire_2018}) 
who termed it as settling instability, 
SI hereafter. Along with the streaming instability, SI is another manifestation of RDI caused by sedimentation
of dust. It was revealed that, as compared with the streaming instability, SI growth rate is much higher for small grains with
stopping time shorter than the Keplerian time, because its maximum value does not depend on grain size.
Furthermore, SI operates at larger length-scales. These and other advantages of SI make it a promising candidate for essential link 
in the chain of physical processes leading to planetesimal formation, see the discussion in Section 9.2 of \citetalias{squire_2018}.


This work is focused on the nature of SI. 
The Lagrangian framework is employed to study the dynamics of small perturbations of gas-dust mixture in two-fluid approximation.
Such a framework is well developed in application to perturbations in single-fluid dynamics. 
In particular, the Lagrangian describing the evolution of perturbations enables to introduce the energy 
of perturbations from the fundamental symmetry of the system with respect to translations in time.
The energy of perturbations is conserved provided that the background flow is stationary, however, it
is not necessarily a positive definite quantity in a moving fluid. The existence of negative energy perturbations
in the flow is well established marker of instability. Indeed, any sink of energy in the system, such as e.g. flux of radiation at 
infinity or viscous damping, provides growth of amplitude of negative energy perturbations. 
Thereby, \citet{friedman-1978b} showed that such perturbations cause the secular instability of rotating stars with respect to gravitational radiation. The energy functional derived in terms of the Lagrangian displacement was referred to as the canonical 
energy, see \citet{friedman-1978}. The concept of the canonical energy has been used in various astrophysical applications
including, for example, stability of relativistic stars, see e.g. \citet{friedman-1998}, stability of magnetic stars, see e.g. 
\citet{andersson-2007}, dynamic tides in stars, see e.g. \citet{ivanov-2010}, \citet{ogilvie-2013}, and many others.


The physical sense of negative energy perturbations is easier to grasp noticing that the total mechanical
energy of the corresponding perturbed flow being itself a positive definite quantity is less than the energy of the background flow.
Probably, for the first time the concept of negative energy in application to space-charge waves in electron beams was 
discussed by L.J. Chu (1951), see the monograph by \citet{briggs-1964} and the book by \citet{pierce-2006}. 
\citet{kadomtsev-1965} were the first to show the existence of negative energy electromagnetic waves in plasma, while the first 
analysis of such possibility for waves in hydrodynamical flows goes back to works by \citet{landahl-1962} and \citet{benjamin-1963}.
A bright exposition on the negative energy waves in context of fluid mechanics and various applications can be found, e.g., in the 
review by \citet{ostrovskiy-1986} and the monograph by \citet{stepanyants-fabrikant-1989}.

One more effect, which provides the growth of amplitude of negative energy wave is coalescence, also termed as coupling, or 
resonant interaction with some other wave having positive energy. 
As soon as the former transfers the energy to the latter, the both waves experience an unbound growth.
By the way, an example of such a coupling of space-charge waves with electromagnetic wave traveling along a helical coil surrounding an electron beam, which is described in the book by \citet{pierce-2006} cited above, 
is quite similar to the subject of this work: roughly speaking, it is enough to replace the electron beam by the streaming of dust, the electromagnetic wave in helical coil by inertial wave (IW) in gas environment, the space-charge wave by a streaming dust wave (SDW) and finally, the spatial growth by the growth over time. A detailed description of the mode coupling and classification of its variants in application to the streaming of charged particles through a plasma can be found in \citet{briggs-1964}.
As for hydrodynamic applications, a notable work was done by \citet{cairns-1979} who discovered that the mode coupling is responsible for 
a classic Kelvin-Helmholtz instability. Two waves existing on a jump of fluid density in a uniform gravitational field acquire 
energy of different signs and couple with each other, as soon as the layers of different density are in a relative motion.

Another example, where the mode coupling is responsible for hydrodynamical instability, is the well-known Papaloizou-Pringle 
instability of uniform angular momentum rotating tori, see \citet{papaloizou-pringle-1984}, \citet{papaloizou-pringle-1985}, \citet{papaloizou-pringle-1987}, \citet{kojima-1986a}, \citet{kojima-1989}, \citet{goldreich-1986} and their citations. 
An exhaustive explanation of Papaloizou-Pringle instability belongs to \citet{glatzel-1987a} and \citet{glatzel-1987b}. 
In a simplified two-dimensional model it was shown that small perturbations grow due to the coupling of surface gravity modes and sound
modes attached to the inner and the outer boundaries of tori.
In the subsequent paper \citet{glatzel-1988} additionally investigated the mode coupling in supersonic plane shear flow.





In this work, the Lagrangian describing the linear dynamics of modes of two-fluid perturbations is introduced 
for the general situation of dust streaming both vertically and radially in protoplanetary disc. The problem
is considered in the terminal velocity approximation (TVA). It is found that the energy of certain 
(neutral) mode of such perturbations akin to SDW can be negative. SDW itself is a trivial mode, which also can have negative energy and
represents perturbations of dust density transported by the bulk stream of dust through the unperturbed gas environment in the absence of dust back reaction on gas.
Physically, this is the bulk streaming of dust that gives rise to a reservoir of negative energy in the form of SDW, 
from which IW can draw the unlimited amount of energy. 
The latter occurs as soon as the phase velocity of SDW is sufficiently close to the phase velocity 
of IW. As the dust back reaction on gas is taken into account, both modes coalesce with each other giving rise 
to the pair of damping and growing coupled modes with strictly equal phase velocities and the vanished energy, thus, the band of SI appears. The strength of the mode coupling is determined by the coupling term in the dispersion equation, 
which is proportional to the mass fraction of dust in gas-dust mixture. The growth rate of SI takes maximum value at resonance 
between IW and SDW, where IW and SDW acquire equal phase velocities. It is proportional to the square root of the coupling term and, 
correspondingly, to the square root of the dust fraction. At the same time, it does not depend on the size of the particles, 
see also \citetalias{squire_2018}.

The analysis shows that generally, there are various types of resonances between IW and SDW, also called 
the mode crossings. Not all of them result in the mode coupling. 
If the energy of SDW is positive, the modes fall into the avoided crossing in the vicinity of resonance, which provides no SI.
That is, the dust back reaction on gas breaks the dispersion curves describing SDW and IW and produces two separate branches 
of neutral modes. At the same time, there are resonances which exist in the high wavenumber limit. 
It turns out that the corresponding mode coupling gives rise to an unbounded growth rate of SI found by \citetalias{squire_2018}.
For sufficiently high dust fraction the unbounded growth rate is proportional to its cube root, which is explained by the
triple coupling of SDW and two low-frequency IW. In addition, the low-frequency IW bond to each other due to the dust 
back reaction on gas and give a quasi-resonant instability of a new type, which may be as strong as SI when the dust settling dominates 
the dust radial drift. 

Finally, it is shown that there is no RDI within TVA when vertical gravity is negligible and dust streams only in the radial direction.
The resonance between SDW and IW existing in this case provides the avoided crossing rather then the mode coupling.
The known streaming instability studied since the work of \citet{youdin-goodman-2005} is found beyond TVA. As was shown by \citetalias{squire_2018}, its growth rate is proportional to the square root of the dust 
fraction, thus, it is also RDI. However, this sort of RDI arises with the account of terms describing the inertia of solids, which 
are of the next order in the particle stopping time. The physical mechanism responsible for streaming instability of
\citet{youdin-goodman-2005}, though resonant in nature, has nothing to do with the mode coupling.

The study starts from general equations for local dynamics of gas-dust mixture in protoplanetary disc and the subsequent
review of TVA. The main part of the paper is focused on the particular model of the perturbed gas-dust mixture constructed in TVA 
for homogeneous background with dust streaming vertically and radially in a disc. 
In the last Section the streaming instability of dust drifting in radial direction is studied beyond TVA.

\section{General equations for local dynamics of gas-dust mixture in a disc}

In order to consider a small patch inside protoplanetary disc,
the local Cartesian coordinates $x,y,z$ are introduced to represent, respectively, radial $r$, azimuthal $\varphi$ and
vertical $z$ directions. The radial distance is measured ourwards with respect to the host star, while the vertical one is
collinear with disc rotation axis. The centre of a reference frame is located at some point $r_0,\varphi_0, z_0>0$ 
above the disc midplane rotating with angular velocity $\Omega_0$ around the host star.
Accordingly, there are $x\equiv r-r_0,\, y\equiv r_0(\varphi-\varphi_0),\, z\to (z-z_0)$ and it is assumed below that 
$\{x,y,z\}\ll h\ll r_0$, where $h$ is the disc scaleheight. This corresponds to the small shearing box approximation, 
see \citet{goldreich-lynden-bell-1965} and the Appendix A of \citet{umurhan-regev-2004} for detailed derivation of fluid equations.
In this work the same approximation is used to address the two-fluid dynamics of gas-dust mixture.
Both gas and dust are considered as fluids with velocities, respectively, ${\bf U}_g$ and ${\bf U}_p$
measured with respect to some reference velocity ${\bf U}_0$. The latter describes the stationary circular shear
motion of a single fluid in the gravitational field of the host star,
\begin{equation}
\label{U_0}
{\bf U}_0 = -q\Omega_0 x\, {\bf e}_y,
\end{equation}
obeying the following radial and vertical balance 
\begin{equation}
\label{rad_0}
\frac{1}{\rho_g}\frac{\partial p_0}{\partial x} = -\frac{\partial \Phi}{\partial x} + \Omega_0^2 (r_0+x) + 2\Omega_0 U_0,
\end{equation}
\begin{equation}
\label{vert_0}
\frac{1}{\rho_g}\frac{\partial p_0}{\partial z} = -\frac{\partial \Phi}{\partial z},
\end{equation}
where $\Phi$ is the Newtonian point-mass gravitational potential, $p_0$ and $\rho_g$ are pressure and density of the fluid, which can be obtained given the equation of state.
In eq. (\ref{U_0}) the shear rate $q$ is assumed to be constant normally close to the Keplerian value $q=3/2$ in protoplanetary discs.
It can be shown, see the Appendix A of \citet{umurhan-regev-2004}, that in the small shearing box approximation the Euler equation for gas reads 
\begin{equation}
\label{eq_U_g}
\begin{aligned}
(\partial_t - q\Omega_0 x \partial_y) {\bf U}_g - 2\Omega_0 U_{g,y} {\bf e}_x + (2-q)\Omega_0 U_{g,x} {\bf e}_y + \\
({\bf U}_g \nabla){\bf U}_g = - \frac{\nabla p}{\rho_g} + \frac{\rho_p}{\rho_g} \frac{\bf V}{t_s},
\end{aligned}
\end{equation}
where $U_{g,x}$ and $U_{g,y}$ are, respectively, $x$- and $y$-projections of ${\bf U}_g$, ${\bf e}_x$ and ${\bf e}_y$ are, 
respectively, $x$- and $y$-orts, while $p$ introduces addition 
to pressure, $p_0$, caused by the influence of dust. The last term in eq. (\ref{eq_U_g}) originates 
from aerodynamic drag of dust particles moving through the gas with the relative velocity ${\bf V}\equiv {\bf U}_p -{\bf U}_g$ 
and mass density $\rho_p$. 
The dust particles stopping time $t_s=const$ is a convenient quantity to parametrise aerodynamic drag, see \citet{whipple-1972}.
On the scale much shorter than $h$ the low-frequency dynamics of gas is vortical, thus, eq. (\ref{eq_U_g}) 
is accompanied by the condition of the divergence-free motion
\begin{equation}
\label{eq_rho}
\nabla \cdot {\bf U}_g = 0.
\end{equation}
Note that in the case of $\rho_p=0$ equations (\ref{eq_U_g}-\ref{eq_rho}) follow from equations (A.28a-d) of \citet{umurhan-regev-2004}
for unstratified fluid, i.e. provided that their $\rho^\prime$=0.

Similarly, the local dynamics of the pressureless fluid, which mimics the solids suspended in gas environment, is described by 
the following equation
\begin{equation}
\label{eq_U_p}
\begin{aligned}
(\partial_t - q\Omega_0 x \partial_y) {\bf U}_p - 2\Omega_0 U_{p,y} {\bf e}_x + (2-q)\Omega_0 U_{p,x} {\bf e}_y + \\
({\bf U}_p\nabla){\bf U}_p = \frac{\nabla p_0}{\rho_g} - \frac{{\bf V}}{t_s},
\end{aligned}
\end{equation}
where $U_{p,x}$ and $U_{p,y}$ are, respectively, $x$- and $y$-projections of ${\bf U}_p$.
The continuity equation for dust reads
\begin{equation}
\label{eq_sigma}
(\partial_t - q\Omega_0 x \partial_y) \rho_p + \nabla (\rho_p{\bf U}_p) = 0.
\end{equation}
Again, eqs. (\ref{eq_U_p}-\ref{eq_sigma}) follow from equations (A.15-A.18) of \citet{umurhan-regev-2004} provided that 
their $P=0$, $\rho\to\rho_p$ and aerodynamic friction is added in its right-hand side (RHS).
As it should be according to the Newton's third law, the aerodynamic drag of dust acting on gas of unit volume equals to the aerodynamic friction of gas acting on dust of unit volume taken with the opposite sign.

In the case when dust is tightly coupled to the gas it is convenient to rewrite equations in terms of the center-of-mass velocity, see \citet{youdin-goodman-2005}
\begin{equation}
\label{def_U}
{\bf U} \equiv \frac{\rho_g{\bf U}_g + \rho_p {\bf U}_p}{\rho},
\end{equation}
where $\rho\equiv \rho_g+\rho_p$ is the total density of gas-dust mixture.

Since
\begin{equation}
\label{U_g_V}
{\bf U}_g = {\bf U} - \frac{\rho_p}{\rho} {\bf V},
\end{equation}
\begin{equation}
\label{U_p_V}
{\bf U}_p = {\bf U} + \frac{\rho_g}{\rho} {\bf V},
\end{equation}
one arrives at the new equations
\\
\\
\begin{equation}
\label{eq_U}
\begin{aligned}
(\partial_t - q\Omega_0 x \partial_y) {\bf U} - 2\Omega_0 U_y {\bf e}_x + (2-q) \Omega_0 U_x {\bf e}_y + ({\bf U}\nabla) {\bf U} + \\ 
\frac{\rho_g}{\rho} \left \{  \left ( {\bf V} \nabla \left ( \frac{\rho_p}{\rho} \right ) \right ) {\bf V} +
2\frac{\rho_p}{\rho} \left ( {\bf V} \nabla \right ) {\bf V} \right \} 
= \frac{\nabla p_0}{\rho_g} - \frac{\nabla(p+p_0)}{\rho},
\end{aligned}
\end{equation}

\begin{equation}
\label{eq_V}
\begin{aligned}
(\partial_t - q\Omega_0 x \partial_y) {\bf V} 
- 2\Omega_0 V_y {\bf e}_x + (2-q) \Omega_0 V_x {\bf e}_y + \\ ({\bf U}\nabla) {\bf V}  + ({\bf V}\nabla) {\bf U} + 
\frac{\rho_g}{\rho} ({\bf V} \nabla) \left ( \frac{\rho_g}{\rho} {\bf V} \right ) - \\ 
\frac{\rho_p}{\rho} ({\bf V} \nabla) \left ( \frac{\rho_p}{\rho} {\bf V} \right ) 
 =  \frac{\nabla (p+p_0)}{\rho_g} - \frac{\rho}{\rho_g} \frac{{\bf V}}{t_s},
\end{aligned}
\end{equation}

\begin{equation}
\label{eq_rho_g}
\nabla \cdot \left ( {\bf U} - \frac{\rho_p}{\rho}{\bf V} \right ) = 0,
\end{equation}

\begin{equation}
\label{eq_rho_tot}
(\partial_t - q\Omega_0 x \partial_y) \rho_p + \nabla ( \rho {\bf U} ) = 0.
\end{equation}


\subsection{Ordering of the terms entering the equations}

Let the characteristic time- and length-scales of gas-dust mixture dynamics be $t_{ev}$ and $l_{ev}$, respectively, while 
the dust mass fraction
\begin{equation}
\label{f}
f\equiv \frac{\rho_p}{\rho_g} < 1.
\end{equation}
Hereafter, the condition (\ref{f}) implies that gas-dust mixture is not dominated by dust.
If the absolute value of the specific pressure gradient, which governs the dynamics of gas-dust mixture, is
$$
g \equiv \left| \frac{\nabla (p + p_0)}{\rho} \right |,
$$
then the restrictions 
\begin{equation}
\label{tva_1}
\tau_* \equiv t_s \max\{ t_{ev}^{-1},\Omega_0 \} \ll 1,
\end{equation}

\begin{equation}
\label{tva_2}
\lambda^{-1} \equiv \frac{g t_s^2}{l_{ev}} \ll 1,
\end{equation}
greatly simplify equations for gas-dust dynamics. 
More exactly, the terms in the left-hand side (LHS) of eq. (\ref{eq_V}) become small compared to 
the terms in its RHS, while the terms $\sim O(V^2)$ become small compared to the rest terms in LHS of eq. (\ref{eq_U}).
This can be justified as follows.
%
%
%
%

If the restrictions (\ref{tva_1}-\ref{tva_2}) are valid, the specific pressure gradient entering the RHS of eq. (\ref{eq_V}) is balanced by the leading term of this equation, which introduces aerodynamic drag. Dividing eq. (\ref{eq_U}) and eq. (\ref{eq_V}) by $g$ one finds that each of these equations consists of the dimensionless terms of various order in the small $\tau_*$ and $\lambda^{-1}$. 
So, the terms in the RHS of eq. (\ref{eq_U}) and eq. (\ref{eq_V}) as well as the gradient term $\sim U^2$ in the LHS of eq. (\ref{eq_U})
are of the zero order in both $\tau_*$ and $\lambda^{-1}$. 
The inertial terms $\sim U$ in the LHS of eq. (\ref{eq_U}) differ from unity by factor $\tau_*/\sqrt{\lambda^{-1}}$. 
Next, the inertial terms $\sim V$ in the LHS of eq. (\ref{eq_V}) 
are of the order of $\tau_*$, whereas the gradient terms $\sim UV$ in the LHS of eq. (\ref{eq_V}) and the gradient
terms $\sim V^2$ in the both of these equations are, respectively, of the order of $\sqrt{\lambda^{-1}}$ and $\lambda^{-1}$. 
Additionally, the following order-of-magnitude relations are valid
$$
V \sim g t_s, \quad U \sim \frac{g t_s} {\sqrt{\lambda^{-1}}},
$$
so that
$$
V/U \sim \sqrt{\lambda^{-1}}.
$$


In order to study the instability of cooperative settling and radial drift of the dust in Section \ref{sec_within_TVA}, 
all terms of the orders of $\tau_*$, $\sqrt{\lambda^{-1}}$ and $\lambda^{-1}$  
in eqs. (\ref{eq_U}) and (\ref{eq_V}) are omitted, 
which corresponds to what is referred to as TVA, 
see \citet{youdin-goodman-2005}. Physically, this means that the inertial forces acting 
on solids in the frame comoving with gas are small compared to drag force and effective gravity force measured in this frame. 
Note that the latter equals to the pressure gradient. As eqs. (\ref{tva_1}) and (\ref{tva_2}) imply, TVA 
is the case of sufficiently small solids with stopping time much shorter than the dynamical time-scale of the problem,
as well as for length-scales of gas-dust mixture dynamics much longer than the solids stopping length $\sim gt_s^2$.

Later on, in Section \ref{sec_beyond_TVA} the full set of eqs. (\ref{eq_U}-\ref{eq_rho_tot}) is reconsidered in order to recover
RDI in the case there is only the radial drift of the dust.

\section{RDI within TVA}

\label{sec_within_TVA}

From now on, eqs. (\ref{eq_U}-\ref{eq_V}) read
\begin{equation}
\label{eq_U_2}
\begin{aligned}
(\partial_t - q\Omega_0 x \partial_y) {\bf U} - 2\Omega_0 U_y {\bf e}_x + (2-q) \Omega_0 U_x {\bf e}_y + ({\bf U}\nabla) {\bf U} = \\ 
\frac{\nabla p_0}{\rho_g} - \frac{\nabla(p+p_0)}{\rho},
\end{aligned}
\end{equation}

\begin{equation}
\label{eq_TVA}
\frac{\nabla (p+p_0)}{\rho} = \frac{{\bf V}}{t_s}.
\end{equation}
This Section deals with the analysis of the particular linear solution of the set of eqs. (\ref{eq_U_2}), (\ref{eq_TVA}), (\ref{eq_rho_g}) and (\ref{eq_rho_tot}). In what follows, where necessary, the usual dimensionless stopping time is used 
$\tau \equiv t_s\Omega_0$.

\subsection{Stationary solution}
\label{gen_stat_sol}

In the reference frame rotating with $\Omega_0$, the local gravitational acceleration has both 
vertical and radial components. In the geometrically thin disc, the former is defined with the help of eq. (\ref{vert_0}),
\begin{equation}
\label{g_z}
-\frac{\partial_z p_0}{\rho_g} \equiv g_z \approx \Omega_0^2 z_0 > 0, 
\end{equation}
and may be considered as a constant value inside the small box. Note that in the last equality in eq. (\ref{g_z}) 
the Keplerian value of angular velocity, $\Omega_K$, is replaced by $\Omega_0<\Omega_K$. 
This is justified by the small difference between $\Omega_0$ and $\Omega_K$, which is of the order of $(h/r_0)^2$.
This difference is usually parametrised by the dimensionless 
$$
\eta \equiv \frac{\Omega_K-\Omega_0}{\Omega_K} \ll 1.
$$
Eq. (\ref{rad_0}) gives the following relation between the radial (effective) gravitational acceleration and $\eta$: 
\begin{equation}
\label{g_x}
-\frac{\partial_x p_0}{\rho_g} \equiv g_x \approx 2\eta \Omega_0^2 r_0 > 0. 
\end{equation}
Depending on the height of the box above the disc midplane, $g_x$ and $g_z$ can be of any ratio with each other.


The most simple solution of eqs. (\ref{eq_U_2}), (\ref{eq_TVA}), (\ref{eq_rho_g}) and (\ref{eq_rho_tot}) 
describing the dust settling is
\begin{equation}
\label{bg_U}
{\bf U} = 0,
\end{equation}
\begin{equation}
\label{bg_p}
\frac{\nabla (p+p_0)}{\rho} = {\bf g},
\end{equation}

\begin{equation}
\label{bg_V}
{\bf V} = t_s {\bf g},
\end{equation}
where ${\bf g} = - g_x {\bf e}_x - g_z {\bf e}_z$.
Hence, the velocity of dust streaming through the gas environment is directed along the (effective) gravitational acceleration
in the box.

Finally, eq. (\ref{eq_rho_g}) along with eq. (\ref{bg_V}) implies that
\begin{equation}
\label{bg_sigma}
\rho_p=const
\end{equation}
on the local scale considered here.

Eqs. (\ref{bg_U}-\ref{bg_sigma}) is the local variant of known \citet{Nakagawa-1986} solution.

\subsection{Equations for axisymmetric perturbations}

The reduced set of equations describing the dynamics of gas-dust perturbations with the background (\ref{bg_U}-\ref{bg_sigma})
in the limit of small $f\ll~1$ is derived in the Appendix \ref{App_1}. The appropriate state variables are
\begin{equation}
\label{Varpi}
\varpi \equiv -\partial_z u_y, \nonumber
\end{equation}
\begin{equation}
\label{Phi}
\phi \equiv \partial_z u_x, \nonumber
\end{equation}
$u_z$ and $\delta$, where $u_x,u_y,u_z$ are the components of the Eulerian perturbation of the centre-of-mass velocity and 
$\delta$ is the relative perturbation of the dust density, see the Appendix \ref{App_1} for the details.

The basic equations take the form:
\begin{equation}
\label{Sys_1}
\partial_t \phi = \partial^2_{tx} u_z  - 2\Omega_0 \varpi + f (g_z \partial_x \delta - g_x \partial_z \delta),  
\end{equation}
\begin{equation}
\label{Sys_2}
\partial_t \varpi = \frac{\kappa^2}{2\Omega_0} \phi,
\end{equation}
\begin{equation}
\label{Sys_3}
\partial^2_{tx}\varpi = - \frac{\kappa^2}{2\Omega_0} \partial^2_{zz} u_z,
\end{equation}
\begin{equation}
\label{Sys_4}
\partial^2_{tz}\delta = t_s (g_z \partial^2_{zz}\delta + g_x \partial^2_{xz} \delta  ) + 2\tau \partial_x \varpi,
\end{equation}
where $\kappa^2 \equiv 2(2-q)\Omega_0^2$ is the epicyclic frequency squared.

Let the gas-dust perturbation have the form of the plane wave
\begin{equation}
\label{fourier}
\chi_i = \hat \chi_i \exp(-{\rm i}\omega t + {\rm i} {\bf k x}).
\end{equation}
That is, $\hat \chi_i$ are complex Fourier amplitudes of $\chi_i$.
In eq. (\ref{fourier}) ${\bf kx } = k_x x + k_z z$, where $k_x$ and $k_z$ are wavenumbers, respectively, along radial and vertical directions in a disc. The frequency $\omega$ is in general a complex value, so that $\Im[\omega]>0$ implies the exponential growth 
of wave amplitude, i.e. the onset of SI.

Eqs. (\ref{Sys_1}-\ref{Sys_4}) yield
\begin{equation}
\label{mode_eq_1}
-{\rm i} \omega \hat \phi - \omega k_x \hat u_z + 2\Omega_0 \hat \varpi - f g_z {\rm i} k_x \hat \delta 
+ f g_x {\rm i} k_z \hat \delta = 0,
\end{equation}
\begin{equation}
\label{mode_eq_2}
-{\rm i} \omega \hat \varpi - \frac{\kappa^2}{2\Omega_0} \hat \phi = 0,
\end{equation}
\begin{equation}
\label{mode_eq_3}
\omega k_x \hat \varpi - \frac{\kappa^2}{2\Omega_0} k_z^2 \hat u_z = 0,
\end{equation}
\begin{equation}
\label{mode_eq_4}
\omega k_z \hat \delta + t_s g_z k_z^2 \hat \delta + t_s g_x k_x k_z \hat \delta - 2\tau {\rm i} k_x \hat \varpi = 0.
\end{equation}

\subsection{Energy of modes of gas-dust perturbations: derivation from variational principle}

For class of neutral modes, when $\Im [\omega]=0$ and the amplitudes are constant, it is possible to formulate 
the variational principle describing the dynamics of small perturbations.

Let the modal solution have the form
\begin{equation}
\label{real_mode}
\begin{aligned}
\varpi = \tilde \varpi \cos \theta, \\
\phi = \tilde \phi \sin \theta, \\
u_z = \tilde u_z \cos \theta, \\
\delta = \tilde \delta \sin \theta, \\
\end{aligned}
\end{equation}
where tilded quantities are real and the phase $\theta \equiv -\omega t + {\bf k x}$. 
%
The tilded quantities in eq. (\ref{real_mode}) satisfy the set of eqs. (\ref{mode_eq_1}-\ref{mode_eq_4}), where
the following replacement is made
$$
\hat \varpi \to \tilde \varpi,\, \hat \phi \to -{\rm i}\tilde \phi,\, \hat u_z \to \tilde u_z, \, \hat \delta \to -{\rm i} \tilde \delta. 
$$

According to the averaged variational principle for modes, see \citet{whitham}, there is a Lagrangian 
$$
L = L(\tilde\chi_i, \partial_i\theta),
$$
where $\tilde\chi_i \equiv \{ \tilde\varpi,\tilde\phi,\tilde u_z,\tilde \delta \}$ and 
$\partial_i \equiv \{ \partial_t, \partial_x, \partial_z \}$, 
such that the corresponding Euler-Lagrange equations are equivalent to the set of eqs. (\ref{mode_eq_1}-\ref{mode_eq_4}) for modes of perturbations provided that the action
\begin{equation}
\label{action_modes}
S = \int { L} d^2{\bf x}\, dt
\end{equation}
with $d^2{\bf x} \equiv dxdz$ be stationary with respect to arbitrary variations of $\tilde \chi_i$ and $\theta$ 
in the small shearing box. The Lagrangian density is
\begin{equation}
\label{L_full_modes}
{L} = {L}_0 + f {L}_1,
\end{equation}
where
\begin{equation}
\label{L_0_modes}
L_0 = \omega \tilde \varpi \tilde \phi + \omega k_x \tilde \varpi \tilde u_z - \Omega_0 \tilde \varpi^2 - 
\frac{\kappa^2}{2\Omega_0} \frac{\tilde \phi^2}{2} -  \frac{\kappa^2}{2\Omega_0} \frac{k_z^2 \tilde u_z^2}{2}
\end{equation}
and
\begin{equation}
\label{L_1_modes}
L_1 = g_z \left [ 1 - \frac{k_z}{k_x} \frac{g_x}{g_z} \right ] 
\left \{ k_x \tilde\varpi\tilde\delta + \frac{k_z \tilde \delta^2}{4\Omega_0} \left [ \frac{\omega}{t_s} + 
(g_z k_z + g_x k_x) \right] \right \}.
\end{equation}

The subscript `0' denotes the contribution to the full Lagrangian responsible for the dynamics of gas-dust perturbations 
with no account for the dust back reaction on gas. In the absence of dust, $L_0$ describes single-fluid vortical perturbations in rotating gas. 
The subscript `1' denotes additional contribution to $L$ responsible solely for the dynamics of dust in terms 
of the evolution of its density. The first term in curly braces in $L_1$ is the cross-term, which describes both the action of gas on solids and 
the back reaction of solids on gas via the aerodynamic drag. 

It can be checked that eqs. (\ref{mode_eq_1}-\ref{mode_eq_4}) correspond to the following Euler-Lagrange equations 
\begin{equation}
\label{aver_L_eq}
\frac{\partial L}{\partial \tilde\chi_i} = 0.
\end{equation}
Also, it is straightforward to check that equations (\ref{aver_L_eq}) yield an alternative form of the dispersion equation 
(cf. eq. (\ref{disp}) below), which is nothing but
\begin{equation}
\label{disp_L}
L=0.
\end{equation}
The remaining Euler-Lagrange equation is
\begin{equation}
\label{aver_L_eq_2}
-\partial_t \left ( \frac{\partial L}{\partial \omega} \right ) + 
\partial_x \left ( \frac{\partial L}{\partial k_x} \right ) + \partial_z \left ( \frac{\partial L}{\partial k_z} \right ) = 0.
\end{equation}

Note that ${L}$ is known up to an arbitrary (dimensional) constant factor since it describes the linear problem.
The sign of ${L}$ is chosen so that the energy of IW propagating in rigidly rotating fluid 
(i.e. in the case $f=q=0$) be positive, see eq. (\ref{E_0_mode}) along with eq. (\ref{disp_inert}) below.

\vspace{0.1cm}

A conserved quantity associated with the invariance of ${L}$ with respect to translations in time is the 
energy of neutral mode
$E \equiv \int {\hat E} d^2{\bf x}$, where
\begin{equation}
\label{E_mode_def}
{\hat E} = \frac{\partial L}{\partial (\partial_t \theta)} \partial_t \theta = \omega \frac{\partial L}{\partial \omega}
\end{equation}
is the energy density of mode. 
Eq. (\ref{aver_L_eq_2}) shows that $E$ changes due to the surface flux of
\begin{equation}
\label{F_mode_def}
{\bf F} \equiv - \omega \frac{\partial L}{\partial {\bf k}}.
\end{equation}
Hence, this quantity is associated with the energy flux of mode, see \citet{whitham}.

The energy density of mode incorporates the basic contribution, 
which is similar to the energy density of gas perturbations in rotating flow in the absence of dust, 
and the additional contribution related to perturbations of dust density
\begin{equation}
\label{E_full_modes}
\hat {E} = \hat {E}_0 + f \hat {E}_1.
\end{equation}
Explicitly,
\begin{equation}
\label{E_0_mode}
\hat E_0 = 2\Omega_0 \frac{k^2 \omega^2}{\kappa^2 k_z^2} \, \tilde \varpi^2,
\end{equation}
where $k^2 = k_x^2 + k_z^2$, and
\begin{equation}
\label{E_1_mode}
\hat E_1 = \frac{\omega}{4\tau} \frac{k_z}{k_x} (k_x g_z - k_z g_x) \, \tilde\delta^2.
\end{equation}

For what follows, see Sections \ref{sec_settling}, \ref{sec_avoided_crossing} and \ref{sec_k_x_neg}, 
it is convenient to represent $\hat E$ in the form:
\begin{equation}
\label{E_mode_varpi}
\hat E = \left \{ 2\Omega_0 \frac{k^2 \omega^2}{\kappa^2 k_z^2} + 
f\tau \omega \frac{k_x g_z - k_z g_x}{(\omega + t_s g_z k_z + t_s g_x k_x)^2} \frac{k_x}{k_z} \right \} \, \tilde\varpi^2.
\end{equation}

In the limit $f\to 0$ $\hat E\to \hat E_0$, which is positive definite\footnote{This is true in centrifugally stable flows only}. 
However, $\hat E_1$ has any sign depending on the signs of wavenumbers and frequency of mode,
as well as on the sign of combination in the brackets of eq. (\ref{E_1_mode}). 
As the dust fraction becomes non-negligible, the balance of terms in eq. (\ref{E_full_modes}) 
for the corresponding mode frequency $\omega(k_x,k_z)$ determines whether $\hat E$ is
either positive or negative quantity.

In the particular case of the dust settling, i.e. when $g_x\to 0$, it is possible to do a straightforward generalisation 
of the above variational principle onto perturbations of general form, see Appendix \ref{App_2}. 
It can be shown that in the general case of dust streaming vertically and radially in a disc the variational principle
of the form (\ref{action}) should contain the higher derivatives of $\chi_i$. The development of such a formalism is beyond
the scope of this study.
Note that in the case $g_x\to 0$  eq. (\ref{E_mode_varpi}) becomes identical to eq. (\ref{E_mode}) provided that $\Im[\omega]=0$. 


\subsection{Dispersion equation}
\label{sec_disp_eq}

As it follows from eqs. (\ref{mode_eq_1}-\ref{mode_eq_4}), $\omega$ obeys an equation 
\begin{equation}
\label{disp}
D_g(\omega,{\bf k}) \cdot D_p(\omega,{\bf k}) = \epsilon({\bf k}),
\end{equation}
where 
\begin{equation}
\label{D_g}
D_g(\omega,{\bf k}) \equiv \omega^2 - \omega_i^2,
\end{equation}
\begin{equation}
\label{D_p}
D_p(\omega,{\bf k}) \equiv \omega - \omega_p,
\end{equation}
\begin{equation}
\label{epsilon}
\epsilon({\bf k}) \equiv f t_s \kappa^2 \frac{k_x}{k} \frac{k_z}{k} (k_x g_z - k_z g_x)
\end{equation}
with $\omega_i\equiv -(k_z/k)\kappa$ and $\omega_p \equiv -t_s g_x k_x - t_s g_z k_z$.

For $f\to 0$, eq. (\ref{disp}) splits into two separate dispersion equations
\begin{equation}
\label{disp_inert}
D_g(\omega_1,{\bf k}) = 0
\end{equation}
and
\begin{equation}
\label{disp_dust}
D_p(\omega_2,{\bf k}) = 0
\end{equation}
describing, respectively, two IW\footnote{Referred to as `epicyclic mode' in \citetalias{squire_2018}} 
propagating in opposite directions and one else mode, which is referred to as SDW. 
Indeed, eq. (\ref{disp_inert}) can be obtained from eqs. (\ref{mode_eq_1}-\ref{mode_eq_4}) 
setting $\hat \delta=0$. Since the dust fraction is negligible, the remaining variables describe perturbations of gas only. 
The set of equations (\ref{mode_eq_1}-\ref{mode_eq_4}) in this particular limit is identical to those 
that describe IW in rigidly rotating fluid, see e.g. \citet{landau-lifshitz-1987}, paragraph 14.
In astrophysical context, one should take into account the more general case of shearing rotating flow and replace 
$2\Omega$, which is the epicyclic frequency for rigid rotation, by $\kappa$, see e.g. \citet{balbus-2003}, paragraph 3.2.3.
For this reason, the modes described by eq. (\ref{disp_inert}) are referred to as the IW below\footnote{Note that according to 
eq. (\ref{pert_V}) such IW produce a non-zero perturbation of relative velocity.}.
On the contrary, eq. (\ref{disp_dust}) follows from eqs. (\ref{mode_eq_1}-\ref{mode_eq_4}) provided that 
$\hat \varpi=\hat \phi = \hat u_z = 0$, which means that gas environment remains unperturbed, whereas arbitrary initial perturbations of dust are transported by the bulk streaming of dust with the velocity (\ref{bg_V}). 
The phase velocity of SDW is equal to the scalar product of the dust velocity and the normal to the wavefront.

Accordingly, as long as the dust fraction is negligible, $\hat E_0$ introduces the energy 
of IW, see eq. (\ref{E_0_mode}), whereas $f\hat E_1$ introduces the energy of SDW, see eq. (\ref{E_1_mode}).


\subsection{The mode coupling}
\label{sec_mode_coupling}

The form of the dispersion equation described in the previous Section has been examined in the dynamics of waves in shear flows, 
see \citet{stepanyants-fabrikant-1998}.
\citet{cairns-1979} suggested that the solution of eq. (\ref{disp}) with small RHS can be interpreted
as the coupling of modes having their own dispersion relations $D_g$ and $D_p$. The RHS of eq. (\ref{disp}) is referred to 
as {\it coupling term}.

The coupling term originates from the product of the last two terms in eq. (\ref{mode_eq_1}) and
the last term in eq. (\ref{mode_eq_4}). In turn, the former comes from the last terms in RHS of eqs. (\ref{u_x}) and (\ref{u_z}), 
which are proportional
to gravitational acceleration in a disc. Physically, this term is the deviation of the weight of solids acting on gas, which
emerges due to perturbation of the dust density. At the same time, the last term in eq. (\ref{mode_eq_4}) comes from the first term 
in RHS of eq. (\ref{delta_2}), which exists as soon as the flow is rotational, see eq. (\ref{nabla_W}). 
Indeed, if it were not for the rotation, the pressure maxima would be absent in the perturbed gas environment. In the linear problem, 
the subsonic pressure maxima, which are necessary both for existence of IW and for dust clumping, 
arise due to the action of Coriolis force. The similar situation was noted by \citet{latter-2011} in context of the streaming instability of \citet{youdin-goodman-2005}, see their discussion about the geostrophic balance. 
However, as it becomes clear from the study of \citetalias{squire_2018}, see also Section \ref{sec_beyond_TVA} of this paper, \citet{latter-2011} considered the non-resonant instability of gas-dust mixture different from RDI.


The coupling causes the deviation of the mode 
frequencies by a small value $\Delta \equiv \omega-\omega_c$ at the {\it mode crossing}. The mode crossing is the point in
the phase space, where the phase velocities of modes are equal to each other, thus, 
eqs. (\ref{disp_inert}-\ref{disp_dust}) are jointly satisfied and
$$
\omega_1 = \omega_2 \equiv \omega_c.
$$
Then,
$$
D_g (\omega) \approx \partial_\omega D_g |_{\omega_c} \Delta, 
$$
$$
D_p (\omega) \approx \partial_\omega D_p |_{\omega_c} \Delta. 
$$
Substitution of these approximations into eq. (\ref{disp}) yields the following solution
\begin{equation}
\label{Delta}
\Delta = \pm \left ( \frac{\epsilon}{\partial_\omega D_g|_{\omega_c} \cdot \partial_\omega D_p|_{\omega_c}} \right )^{1/2} = 
\pm \left ( \frac{\epsilon}{2\omega_c} \right)^{1/2}.
\end{equation}
This general result is employed in Sections \ref{sec_settling}-\ref{sec_avoided_crossing} to show a simple way to reproduce 
the analytical estimations of SI growth rate obtained by \citetalias{squire_2018}.

A more subtle thing is that the sign of the expression under the square root in eq. (\ref{Delta})
follows the sign of SDW energy taken at the mode crossing, see eq. (\ref{E_1_mode}). 
Therefore, as there is a mode crossing between SDW and IW, then {\it SI occurs if and only if SDW has negative energy}.
Since the energy of IW is always positive, see eq. (\ref{E_0_mode}), this implies that the instability occurs as soon as 
the coupling waves have the energy of different signs. Following \citet{cairns-1979},
this allows for clear interpretation of SI as the resonant coalescence of waves in the gas-dust mixture, 
which provides an exchange with energy between the waves. The total energy of coupled mode formed by the coalesced waves 
remains constant, while the non-zero energy flux between the coalesced waves provides the growth of their amplitudes 
as soon as the energy flows from the negative energy wave to the positive energy wave. 
On the contrary, if the energy flows in the opposite direction, a coupled mode is damping.
These cases represent the pair of complex conjugate solutions of the dispersion equation (\ref{disp}) 
approximately given by eq. (\ref{Delta}).

The following Sections provide the details of the mode coupling during the dust settling or/and the dust radial drift.
The frequencies and the energy of modes of gas-dust perturbations for finite non-zero $f>0$ 
are obtained from the accurate solution of eq. (\ref{disp}).
Hereafter it is assumed that $k_z>0$. Eq. (\ref{disp}) indicates that the case $k_z<0$ is reduced to the previous one by the
replacement $k_x \to -k_x$ and $\omega \to -\omega$.

\subsection{Units}

Wherever it is needed below, the frequencies are measured in units of $\Omega_0$ and the wavenumbers are measured 
in units of the inversed artificial length-scale, $L\lesssim h$. 
Accordingly, the gravitational acceleration is measured in units of $L\Omega_0^2$. 
In all particular calculations $q=3/2$ and $\tau=0.1$.

\subsection{Particular case of the dust vertical settling}
\label{sec_settling}

Here it is assumed that the box is located sufficiently far above the disc midplane so that $g_z$ solely controls 
the streaming of dust, cf. eqs. (\ref{g_z}) and (\ref{g_x}). The corresponding restriction reads
\begin{equation}
\label{cond_z_0}
\frac{z_0}{r_0} \gg 2\eta.
\end{equation}

\begin{figure}
\begin{center}
\includegraphics[width=8cm,angle=0]{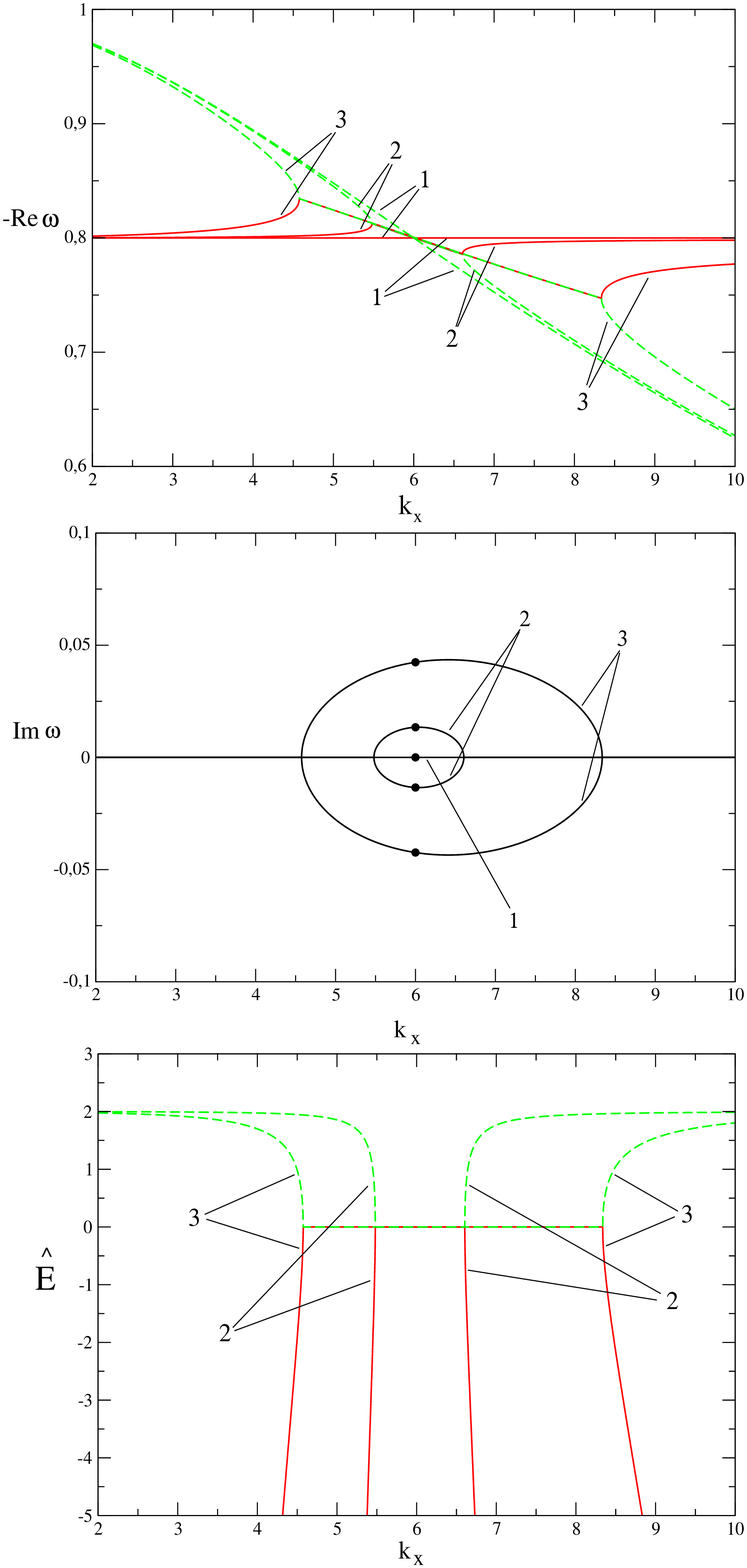}
\end{center}
\caption{
Top panel: the curves show $\Re[\omega]$ taken with the opposite sign, where $\omega$ is the solution of the dispersion equation (\ref{disp}). The parameters are $q=3/2$, $g_x=0$, $g_z=1$, $\tau=0.1$, $k_z=8$. It is assumed that $f=0$ (branches marked as '1'), 
$f=0.001$ (branches marked as '2') and $f=0.01$ (branches marked as '3'). In the case $f=0$ the solid and the dashed line correspond to
SDW and IW (see text), respectively.
In the case $f>0$ the solid and the dashed lines correspond to modes akin to SDW and IW, respectively.
Middle panel: the imaginary part of the corresponding solution.
The filled circles represent the analytical solution given by eq. (\ref{Delta_2}) at the mode crossing given by eq. (\ref{crossing}).
Bottom panel: the energy density of modes with $\Im[\omega]=0$ as given by eq. (\ref{E_mode_varpi}) and 
additionally the energy density of growing/damping modes as given by eq. (\ref{E_mode}), 
for $|\varpi|^2=1$ and the non-zero $f$. 
The solid and the dashed lines correspond to modes akin to SDW and IW, respectively.
} \label{fig_1}
\end{figure}

An accurate solution of eq. (\ref{disp}) for particular values of parameters is presented in Fig. \ref{fig_1}.
The problem is symmetric with respect to $k_x \to -k_x$, so the case $k_x>0$ is considered only.
Note that only two branches of $\omega$ with $\Re [\omega] < 0$ are shown, while there exists a third one with $\Re [\omega] > 0$.
The corresponding profile of the energy density of mode, $\hat E$, normalised by $|\varpi|^2$ 
is shown on the bottom panel of Fig. \ref{fig_1}.

As far as the dust fraction is negligible, $f=0$, there is no instability of gas-dust mixture.
As expected, for $f>0$ SI sets on in the vicinity of the point, where the frequencies of two branches evaluated
for $f=0$ are equal to each other. This is the mode crossing point. 
For the particular values of parameters taken in Fig. \ref{fig_1} it is located at $k_x=6$. 
The non-negligible dust fraction causes a small deviation of 
the branches corresponding to IW and SDW in such a way that in some band around the mode crossing the waves acquire identical 
phase velocities and increment/decrement, which takes maximum value approximately at the mode crossing, compare 
the curves marked by (1) with the curves marked by (2) and (3) in Fig. \ref{fig_1}. 
Strictly speaking, the waves represented by curves (2) and (3) are the specific kind of waves propagating 
through the gas-dust mixture with dust settling to the disc midplane. Outside the band of instability these waves are akin to IW and SDW 
having, respectively, positive and negative energy densities.
As one approaches the band of instability, the absolute value of the energy density of the both branches vanishes. 
At the same time, the neutral waves coalesce and give birth to a pair of 
growing and damping waves as seen in Fig. \ref{fig_1}. 
The energy of such waves vanishes, since this is the only way for the energy
to be conserved for alternating amplitudes of waves, see the Appendix \ref{App2_2} and also \citet{friedman-1978}.
Similar picture occurs in some applications from physics 
of hydrodynamical instability of shear flows, where it is regarded as a mode coupling discussed above 
in Section \ref{sec_mode_coupling}.

According to eqs. (\ref{disp_inert}-\ref{disp_dust}) with $g_x=0$ the mode crossing occurs
as soon as the condition
\begin{equation}
\label{crossing}
t_s g_z = \frac{\kappa}{k}
\end{equation}
is true. Once (\ref{crossing}) is satisfied, the phase velocities of IW and SDW are equal to each other
$\omega_c = \omega_i = \omega_p$.

As one follows derivation of eq. (\ref{crossing}) starting from the general equations (\ref{pert_U}-\ref{pert_rho_tot}), it becomes 
clear that the LHS of eq. (\ref{crossing}) is nothing but the bulk settling velocity of dust given by eq. (\ref{bg_V}), 
also cf. eq. (4.4) of \citet{squire_2018}.
Therefore, the latter defines the characteristic wavelength of the mode coupling and the band of SI.
As has been already recognised by \citet{squire_2018}, the growing gas-dust perturbations of SI have larger length-scales
than that of the streaming instability of \citet{youdin-goodman-2005}. 
Clearly, this is because in the geometrically thin disc settling velocity can be higher than the velocity of radial drift 
by factor $\sim (h/r_0)^{-1}$.

Equation (\ref{Delta}) applied to the case $g_x\to 0$ yields
\begin{equation}
\label{Delta_2}
\Delta = \pm {\rm i} \, \kappa \left ( \frac{f}{2} \right )^{1/2} \frac{k_x}{k},
\end{equation}
where eq. (\ref{crossing}) is taken into account.
The estimate (\ref{Delta_2}) is in accordance with eq. (5.13) of \citetalias{squire_2018}. 
For small value of $f$ used in Fig. \ref{fig_1} this analytical increment/decrement virtually coincides with an 
accurate solution of eq. (\ref{disp}). It is also equivalent to eq. (\ref{approx_im}) in the leading order in $f^{1/2}$.


Additionally, it should be noted that as far as the main restriction of solids tightly coupled to gas is valid, see eq. (\ref{tva_1}),
the condition (\ref{crossing}) fits well within the second restriction of the model, see eq. (\ref{tva_2}).
Indeed, in the case of the Keplerian shear eq. (\ref{crossing}) implies that
$$
l_{ev} \sim g_z t_s \Omega_0^{-1},
$$
which corresponds to the length-scale larger than the stopping length of solids settling to the disc midplane\footnote{cf. 
similar restriction obtained by \citet{latter-2011} who considered the streaming instability.} 
$\sim g_z t_s^2$ by factor $\sim \tau^{-1} \gg 1$.
The  main restriction (\ref{tva_1}) is valid as far as $\Re[\omega] \sim \Omega_0$ and $\Im[\omega] \lesssim \Omega_0$.

\subsection{Introducing the dust radial drift}

\begin{figure}
\begin{center}
\includegraphics[width=8cm,angle=0]{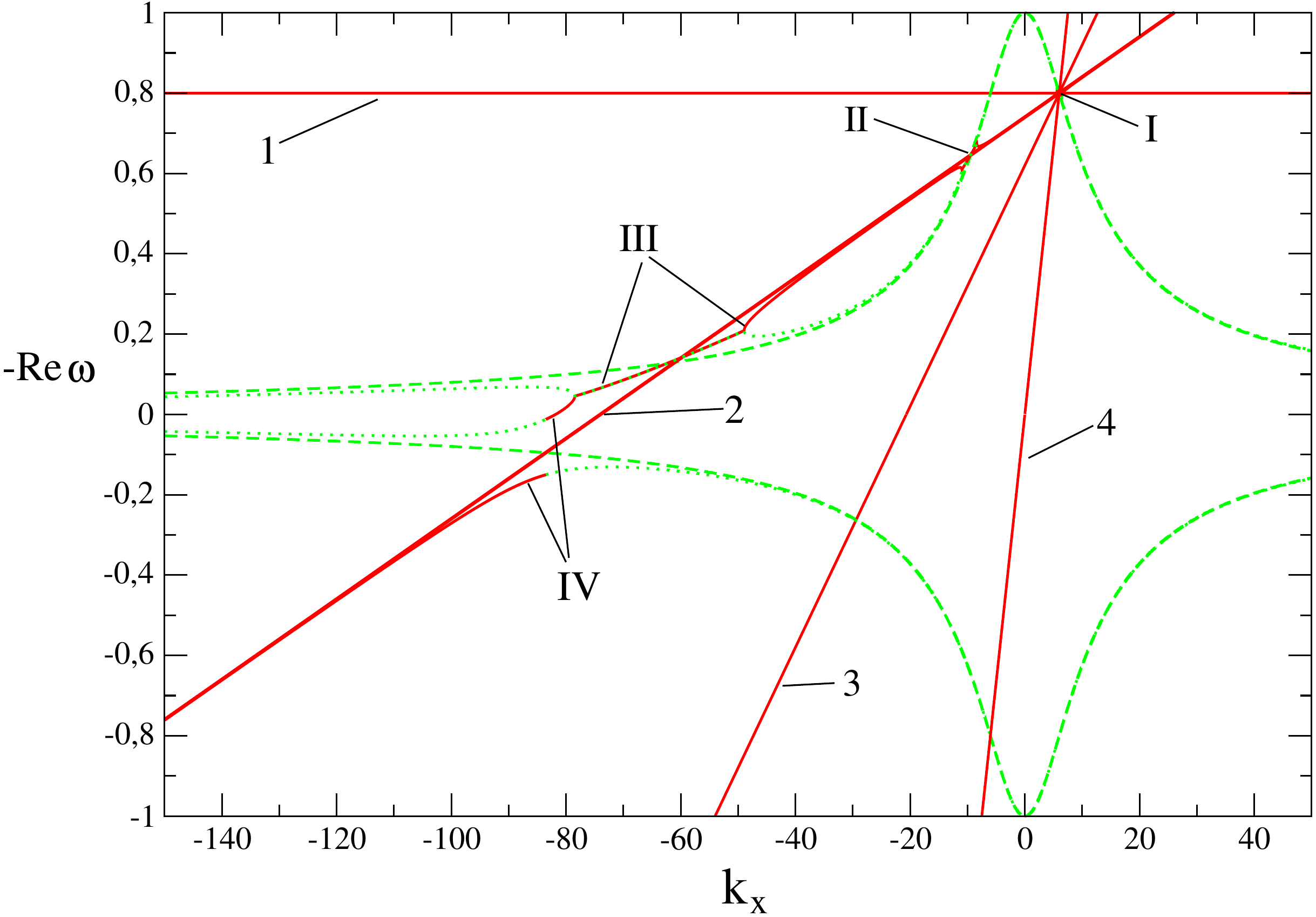}
\end{center}
\caption{
The general view of the mode crossing with the account of the dust radial drift for $\tau=0.1$, $q=3/2$ and $k_z=8$. 
Solid curves marked as `1', `2', `3' and `4' represent SDW for $f=0$ and the value of gravitational acceleration $(g_x;g_z)$ being,
respectively, $(0.0;1.0)$, $(0.1;0.925)$, $(0.3;0.775)$ and $(4/3;0)$, respectively.
Dashed curves represent two IW propagating in the opposite directions. 
Additionally, the solution of the dispersion equation (\ref{disp}) is shown for the case of ($g_x=0.1$; $g_z=0.925$) and $f=0.001$,
see the solid curves marked by `I' `II' `III' and `IV' representing the modes akin to SDW, 
whereas the dotted curves represent the modes akin to IW in this case.
At the same time, these Roman numbers denote the type of the mode crossing which causes the corresponding mode coupling (or avoided crossing, see Sections \ref{sec_avoided_crossing} and \ref{sec_k_x_neg} for details), see table \ref{types}.} 
\label{fig_2}
\end{figure}

As the box gets down to the disc midplane, the bulk streaming of dust is determined by combination of vertical and radial 
components of gravitational acceleration. The symmetry of the problem with respect to $k_x\to -k_x$ breaks and the general picture 
of the mode crossing (just `crossing' below in this Section) between IW and SDW becomes more complex. It is shown in Fig. (\ref{fig_2}) on the plane of $(k_x, -\Re[\omega])$.
For the specific value of $k_z$ there are two certain branches of the oppositely propagating IW given by eq. (\ref{disp_inert}), 
whereas the position of the straight line representing SDW, see eq. (\ref{disp_dust}), is determined by the changeable $g_x$ and $g_z$.
This line is horisontal for $g_x=0$, while it crosses the origin of coordinates, i.e. ($k_x=0,-\Re[\omega]=0$), for $g_z=0$, 
see the lines (1) and (4) in Fig. \ref{fig_2}, respectively. 
In the former limiting case, there can be up to two crossings, whereas in the latter limiting case
there are always two crossings, each one with its own branch of IW.
In the general situation of $g_x>0$ and $g_z>0$, there can be from two and up to the four crossings, see the lines (2) and (3)
in Fig. \ref{fig_2}. Note that the lines of SDW shown in Fig. \ref{fig_2} intersect each other at the crossing
considered for example in Fig. \ref{fig_1}. 

The condition of the crossing becomes
\begin{equation}
\label{gen_crossing}
t_s ( k_x g_x + k_z g_z ) = \pm \kappa \frac{k_z}{k},
\end{equation}
where `+' and `-' are taken for $\omega_c<0$ and $\omega_c>0$, respectively.

In order to facilitate the understanding, it is appropriate to introduce the formal classification of the crossings, 
see table \ref{types} and Fig. \ref{fig_2}. The conditions for existence of the crossings of different types is collected in
the table \ref{existence}.

\begin{table}
\caption{Formal types of the mode crossing for any $g_z>0$}
\label{types}
\begin{tabular}{cccc}
type       & $(k_x,Re[\omega])$ sector   & group velocities & concavity \\ 
\hline
I       & $k_x\geq 0, \omega_c<0$   & $\partial_{k_x} \omega_p \leq 0 < \partial_{k_x} \omega_i$   & any value  \\
II      & $k_x<0, \omega_c<0$   &  $\partial_{k_x} \omega_i < \partial_{k_x} \omega_p \leq 0$   & any value \\
III     & $k_x<0, \omega_c<0$   &  $\partial_{k_x} \omega_p \leq \partial_{k_x} \omega_i<0$   & $\partial^2_{k_x}\omega_i<0$  \\
\ \ III$^\prime$  & $k_x<0, \omega_c<0$   &  $\partial_{k_x} \omega_p \leq \partial_{k_x} \omega_i<0$   & $\partial^2_{k_x}\omega_i>0$  \\
\ IV      & $k_x<0, \omega_c>0$   &  $\partial_{k_x} \omega_p < 0 < \partial_{k_x} (-\omega_i)$   & any value \\
\hline
\end{tabular}
\end{table}

\begin{table}
\caption{Existence of the mode crossings of different types for any $g_z>0$.}
\label{existence}
\begin{tabular}{ll}
type       & existence \\ 
\hline
I       & $\forall g_x$: $k_z \leq \tilde k_z$   \\
II      & $0\leq g_x < g_x^{'}$: $k_z^{'} < k_z < k_z^{''}$  \\
III     & $0<g_x\leq g_x^{'}$: $k_z^{'}<k_z$,\ \ $g_x^{'} < g_x < g_x^{''}$: $\tilde k_z^{'}< k_z$  \\
III$^\prime$  & $0<g_x \leq  g_x^{'}$: $\tilde k_z < k_z \leq k_z^{''}$,\ \ 
$g_x^{'} < g_x \leq g_x^{''}$: $ \tilde k_z < k_z \leq \tilde k_z^{'}$,\\ 
\ \ & $g_x^{''} < g_x$: $\tilde k_z < k_z$  \\
\ IV      & $\forall k_z$: $g_x>0$ \\
\hline
\end{tabular}
\end{table}

There are a few quantities that delimit the whole space of parameters with respect to the existence of crossings of various type.
First, the value of $g_x = g_x^{'}$, which restricts the existence of type-II crossings, is defined by the 
condition that the lines of SDW and IW do not touch each other for $g_x > g_x^{'}$.
It reads
\begin{equation}
\label{g_x_cr}
g_x^{'} \equiv \frac{g_z}{\sqrt{8}}.
\end{equation}

As soon as $g_x < g_x^{'}$, there are two values of $k_z$, $k_z^{''} > k_z^{'}$, 
corresponding to the positions of SDW line tangent to IW curve.
It is straightforward to obtain them for given $g_z$ and $g_x < g_x^{'}$, 
though the corresponding expressions are omitted here to save space. 
Note that in the limit $g_x \to g_x^{'}$ 
\begin{equation}
\label{k_z_cr}
k_z^{'} \to k_z^{''} \to \bar k_z \equiv \frac{4}{3} \sqrt{\frac{2}{3}} \tilde k_z \approx \tilde k_z,
\end{equation}
where
\begin{equation}
\label{k_z_g_z}
\tilde k_z \equiv \frac{\kappa}{t_s g_z}.
\end{equation}
For $k_z = \tilde k_z$ the crossing occurs at $k_x=0$, where the `negative' branch of IW attains its minimum. 
Oppositely, as $g_x\to 0$ $k_x^{'}\to 0$ and $k_x^{''}\to \tilde k_z$.

Next, for a higher $g_x^{'} < g_x < g_x^{''}$, where
\begin{equation}
\label{g_x_cc}
g_x^{''} \equiv \sqrt{2} g_z,
\end{equation}
type-III and type-III' crossings co-exist in the range $k_z > \tilde k_z$ with
the transition into each other at
\begin{equation}
\label{k_z_cc}
\tilde k_z^{'} \equiv \frac{2 \tilde k_z}{\sqrt{3}(\sqrt{2} - g_x/g_z)}.
\end{equation}
As $g_x \to g_x^{''}$, $\tilde k_z^{'} \to \infty$ and type-III crossing disappears for all $k_z$.
Also, note that in the limiting case $g_x = g_x^{'}$ 
$$
\tilde k_z^{'} = \bar k_z.
$$
A caveat should be done, that in this study of SI the distinction between 
type-III and type-III' crossings looks completely formal. Still, it is shown here for the sake of rigorous analysis.

The formal picture presented above leads to the following summary:
\begin{itemize}
\item
In the case of small radial drift, $0<g_x\ll g_z$, type-I (or type-III') and type-II crossings are confined in the
range $k_z \lesssim \tilde k_z$, while type-III crossing exists for unlimited $k_z$. 
For type-III crossing $|k_x| \gg k_z$. 
\item
As radial drift and vertical settling are comparable to each other, $g_x \sim g_z$, type-II crossing disappears, 
while the whole range of $k_z$ is shared between type-I and type-III(') crossings 
for $k_z \leq \tilde k_z$ and $k_z>\tilde k_z$, respectively.
\item
In the case of dominating radial drift, $g_x\gg g_z$, the whole range of $k_z$ is shared between type-I and type-III$^\prime$ 
crossings, while the zone of type-III$^\prime$ crossing shifts to high $k_z$ as $\tilde k_z$ increases with decreasing $g_z$. 
For type-III$^\prime$ crossing $|k_x| \ll k_z$. 
\item
Type-IV crossing exists in all mentioned cases.
\end{itemize}

Prediction of SI for the crossing of any type can be done using the rule revealed in Section \ref{sec_mode_coupling}.
Examination of the energy density of SDW, see eq. (\ref{E_1_mode}), leads to the conclusion that type-I crossing produces SI 
for $k_x/k_z > g_x/g_z$ only. In the opposite case, the mode coupling is absent. Clearly, the greater 
the radial drift, the more severe becomes this restriction on SI, see Section \ref{sec_avoided_crossing} for details.
On the contrary, type-II and type-III(') crossings produce SI with no regards to the parameters, whereas, by the same reasoning,
at the type-IV crossing the mode coupling is always absent, see Section \ref{sec_k_x_neg}.

According to the general formula (\ref{Delta}), the approximate growth/damping rate at the crossing of any type producing SI is
\begin{equation}
\label{Delta_gen}
\Delta = \pm {\rm i} \kappa \left ( \frac{f}{2} \right )^{1/2} \frac{(k_x k_z)^{1/2}}{k} 
\left ( \frac{k_x g_z - k_z g_x}{g_x k_x + g_z k_z} \right )^{1/2}, 
\end{equation}
which must be evaluated under the condition (\ref{gen_crossing}). 
In the case of the crossing, which does not produce SI, eq. (\ref{Delta_gen}) provides an estimation for real correction to $\omega_c$.
Eq. (\ref{Delta_gen}) is in accordance with  eq. (5.12) of \citetalias{squire_2018}.

\subsection{Avoided crossing of modes}
\label{sec_avoided_crossing}

In this Section type-I crossing is considered more closely.
It can be seen that for $g_x>0$ the coupling term given by eq. (\ref{epsilon}) changes its sign at some $k_x>0$.
For smaller $k_x$, the mode coupling ceases even for non-zero dust fraction and is replaced by the {\it avoided crossing}, 
see e.g. \citet{stepanyants-fabrikant-1989}.

The condition $\epsilon=0$ along with the mode crossing condition (\ref{gen_crossing}) yields that for the given $g_x$ and $g_z$
the mode coupling ceases at
\begin{equation}
\label{avoided_k_x}
k_x^{\rm (AC)} = \frac{g_x g_z^2}{g^3} \, \tilde k_z,
\end{equation}

\begin{equation}
\label{avoided_k_z}
k_z^{\rm (AC)} = \frac{g_z^3}{g^3} \, \tilde k_z.
\end{equation}
Accordingly, for $k_z<k_z^{(AC)}$ the type-I crossing causes the mode coupling, while for $k_z \geq k_z^{(AC)}$ it
causes the avoided crossing. As $g_x$ becomes greater, $k_z^{\rm (AC)}$ steeply gets smaller than the 
characteristic $k_z = \tilde k_z$. Examination of eq. (\ref{Delta_gen}) for this case in the limit $g_x \gg g_z$, which
corresponds to $k_x \gg k_z$ for type-I crossing, leads to conclusion that the growth rate of SI decreases like
\begin{equation}
\label{avoided_Delta}
\Im[\omega] \sim O\left ( \frac{g_z}{g_x} \right ) \Im[\omega]_{\rm stl},
\end{equation}
where $\Im[\omega]_{\rm stl} = \kappa \, (f/2)^{1/2}$ is the maximum growth rate given by eq. (\ref{Delta_2}).

In the absence of the dust settling, i.e. for $g_z=0$, there is no coupling for type-I crossing, and 
one arrives at the following estimation for the real corrections to the frequencies of SDW and IW:
\begin{equation}
\label{Delta_real}
\Delta = \pm \kappa \left ( \frac{f}{2} \right )^{1/2} \frac{k_z}{k},
\end{equation}
which is in accordance with eq. (5.10) of \citetalias{squire_2018} to the leading order in $\tau$.

\begin{figure}
\begin{center}
\includegraphics[width=8cm,angle=0]{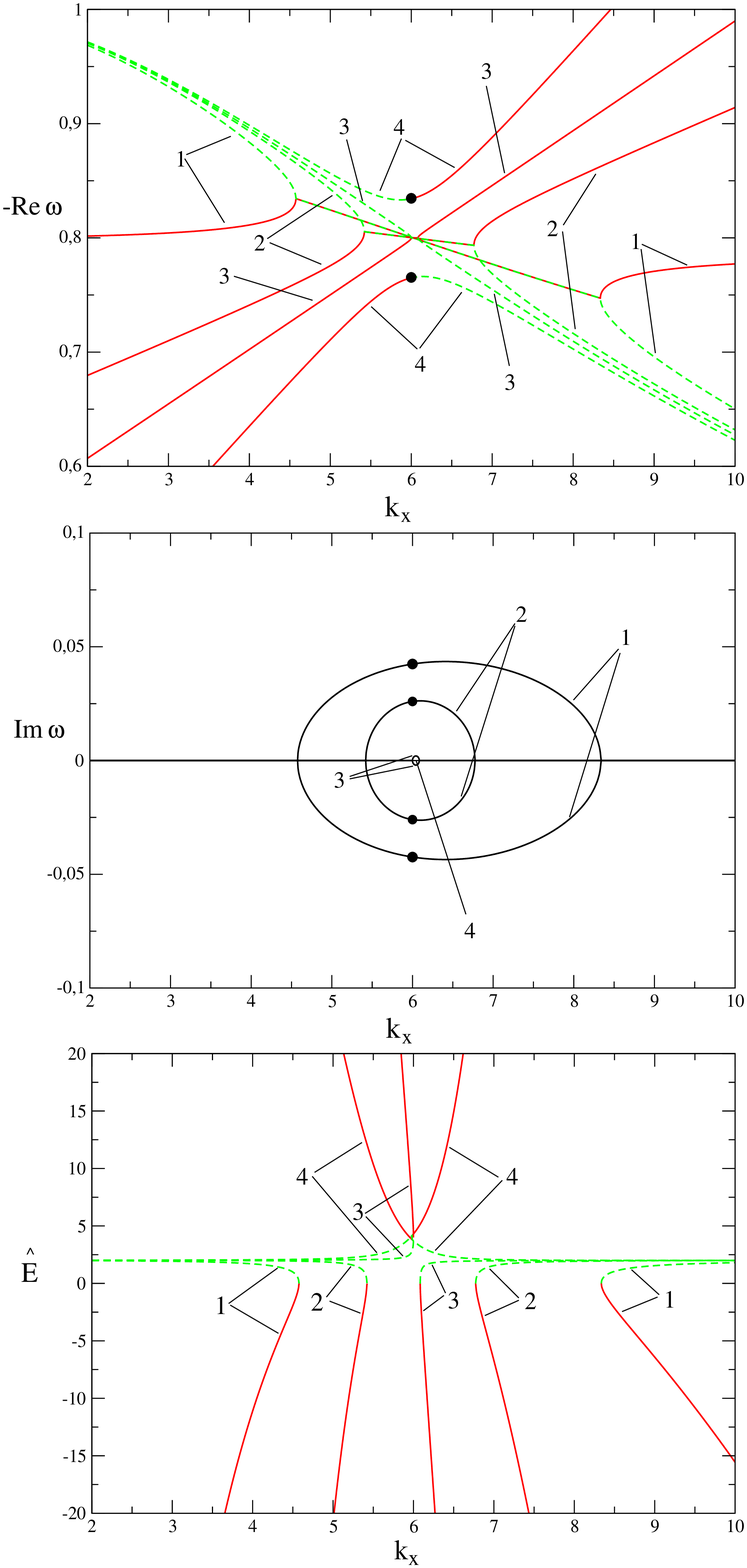}
\end{center}
\caption{Modification of the mode coupling demonstrated in Fig. \ref{fig_1} due to the dust radial drift.
Top panel: the curves show $\Re[\omega]$ taken with the opposite sign, where $\omega$ is the solution of the dispersion equation (\ref{disp}). The parameters are $q=3/2$, $\tau=0.1$, $k_z=8$ and $f=0.01$. 
Gravitational acceleration $(g_x;g_z)$ is the following: $(0.0;1.0)$ (branches marked as '1'), $(0.3;0.775)$ (branches marked as '2'), 
$(0.48;0.64)$ (branches marked as '3') and $(0.8;0.4)$ (branches marked as '4'). 
The solid and the dashed lines correspond to modes akin to SDW and IW, respectively. 
The filled circles represent the analytical solution given by eq. (\ref{Delta_gen}) taken at the mode crossing 
according to eq. (\ref{gen_crossing}) for $(g_x=0.8;g_z=0.4)$.
Middle panel: the imaginary part of the corresponding solution.
The filled circles represent the analytical solution given by eq. (\ref{Delta_gen}) taken at the mode crossing given 
by eq. (\ref{gen_crossing}) for $(g_x=0.0;g_z=1.0)$ and $(g_x=0.3;g_z=0.775)$.
Bottom panel: the energy density of modes as given by eq. (\ref{E_mode_varpi})  for $|\varpi|^2=1$ for modes with 
$\Im[\omega]=0$. 
The solid and the dashed lines correspond to modes akin to SDW and IW, respectively. 
}
\label{fig_3}
\end{figure}

The transition of type-I crossing from the mode coupling to the avoided crossing as the radial drift of the dust becomes
comparable to its settling is shown in Fig. \ref{fig_3}. 
For several particular solutions of eq. (\ref{disp}) the components of gravitational acceleration are chosen in such a way that 
the mode crossing occurs at the same point as in Fig. \ref{fig_1}. It is seen that as $g_x$ increases, the band of SI 
shrinks and the maximum growth rate becomes smaller. Next, for $g_x/g_z \approx 3/4$, which is determined by 
eqs. (\ref{avoided_k_x}-\ref{avoided_k_z}) for this mode crossing, the mode coupling ceases and 
the two separate curves of $\Re[\omega]$ emerge instead. Each of these curves consists of the parts, 
which can be associated with the modes akin to IW and SDW as demonstrated in Fig. \ref{fig_3}.  
In other words, the branches representing the modes 
akin to IW and SDW become broken as the avoided crossing occurs. The corresponding analytical estimations 
done with the account of both the radial drift and the vertical settling are in a good agreement with an accurate result.
Turning to the profiles of the energy of modes, one finds that as far as SI exists, they are similar those shown in Fig. \ref{fig_1}.
However, in the vicinity of the critical point defined by eqs. (\ref{avoided_k_x}-\ref{avoided_k_z}) there is another situation.
Indeed, in the case marked as `3' in Fig. \ref{fig_3} the mode coupling still occurs, though it is rather weak. 
Nevertheless, one of the corresponding branches of mode akin to SDW acquires the positive energy. Further, as the avoided crossing 
arises, one finds that the two separate curves of $\Re[\omega]$ marked as `4' represent the positive energy modes.

\subsection{Behaviour of mode crossings at negative radial wavenumber}
\label{sec_k_x_neg}

All crossings except the type-I crossing take place at $k_x<0$. 
Eq. (\ref{Delta_gen}), or equivalently, the energy density of SDW (\ref{E_1_mode}) along with the rule noted 
in Section \ref{sec_mode_coupling}, indicate that type-II and type-III($^\prime$) crossings result 
in the mode coupling for any $f>0$, while type-IV crossing always provides the real correction to the frequency, thus, 
resulting in the avoided crossing. As follows from table \ref{existence}, 
type-II and type-III crossings occur for the substantial settling only, as they exist for $g_x < g_x^{'}$ and 
$g_x<g_x^{''}$, respectively. At the same time, type-III$^\prime$ and type-IV crossings remain in the limit $g_z\to 0$. 
Additionally, unlike the type-II crossing, type-III,III$^\prime$ and -IV crossings are not confined over $k_z$ and $|k_x|$.

\begin{figure}
\begin{center}
\includegraphics[width=8cm,angle=0]{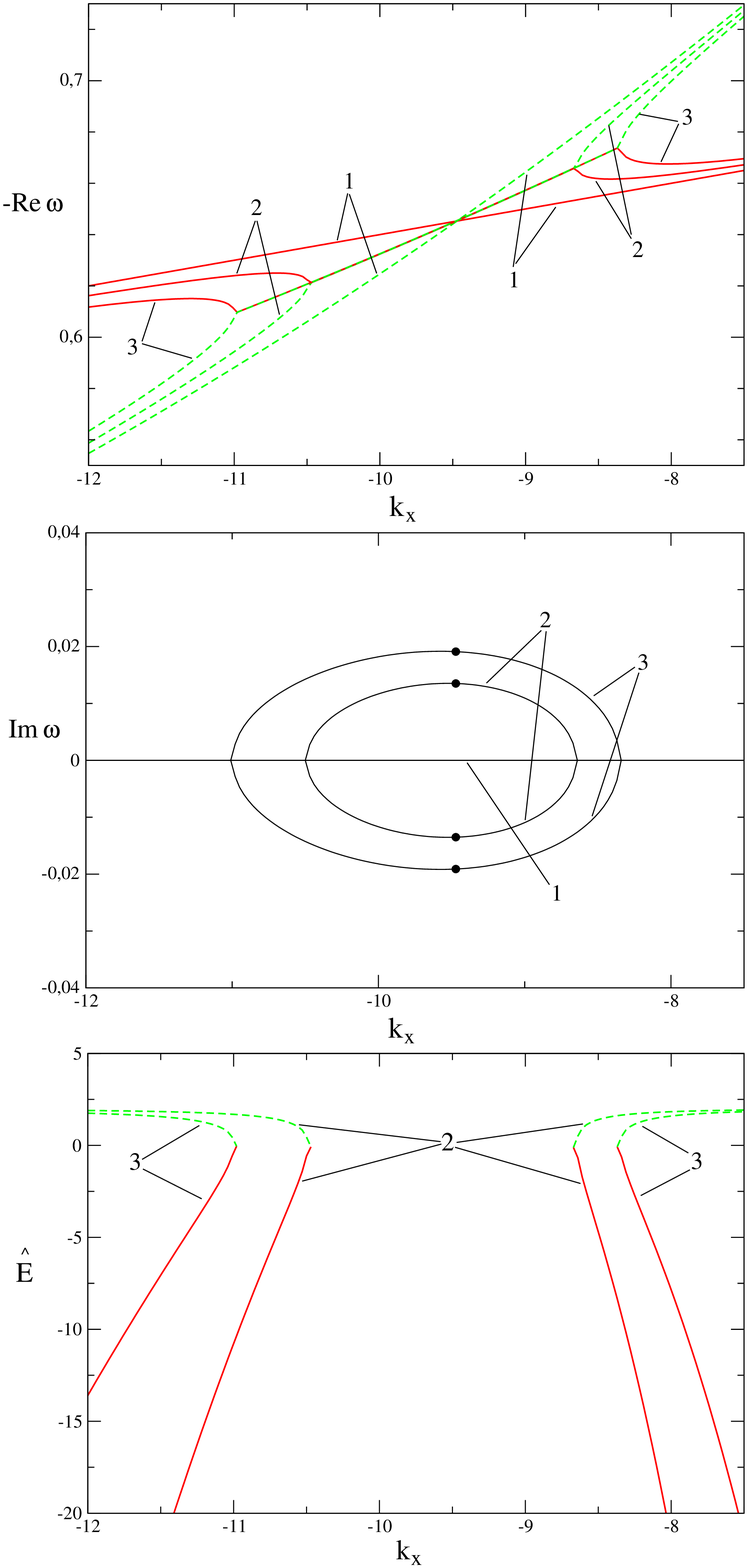}
\end{center}
\caption{The mode coupling produced by the type-II mode crossing. 
Top panel: the curves show $\Re[\omega]$ taken with the opposite sign, where $\omega$ is the solution of the dispersion equation (\ref{disp}). The parameters are $q=3/2$, $\tau=0.1$, $g_x=0.1$, $g_z=0.925$ and $k_z=8$. 
The dust fraction is taken as $f=0$ (branches marked as '1'), 
$f=0.0005$ (branches marked as '2') and $f=0.001$ (branches marked as '3'). 
In the case $f=0$ the solid and the dashed line correspond to SDW and IW, respectively.
In the case $f>0$ the solid and the dashed lines correspond to modes akin to SDW and IW, respectively.
Middle panel: the imaginary part of the corresponding solution.
The filled circles represent the analytical solution given by eq. (\ref{Delta_gen}) at the mode crossing 
according to eq. (\ref{gen_crossing}).
Bottom panel: the averaged energy density of modes with $\Im[\omega]=0$ as given by eq. (\ref{E_mode_varpi})  for $|\varpi|^2=1$ 
and the non-zero $f$. 
The solid and the dashed lines correspond to modes akin to SDW and IW, respectively.
}
\label{fig_4}
\end{figure}

\begin{figure}
\begin{center}
\includegraphics[width=8cm,angle=0]{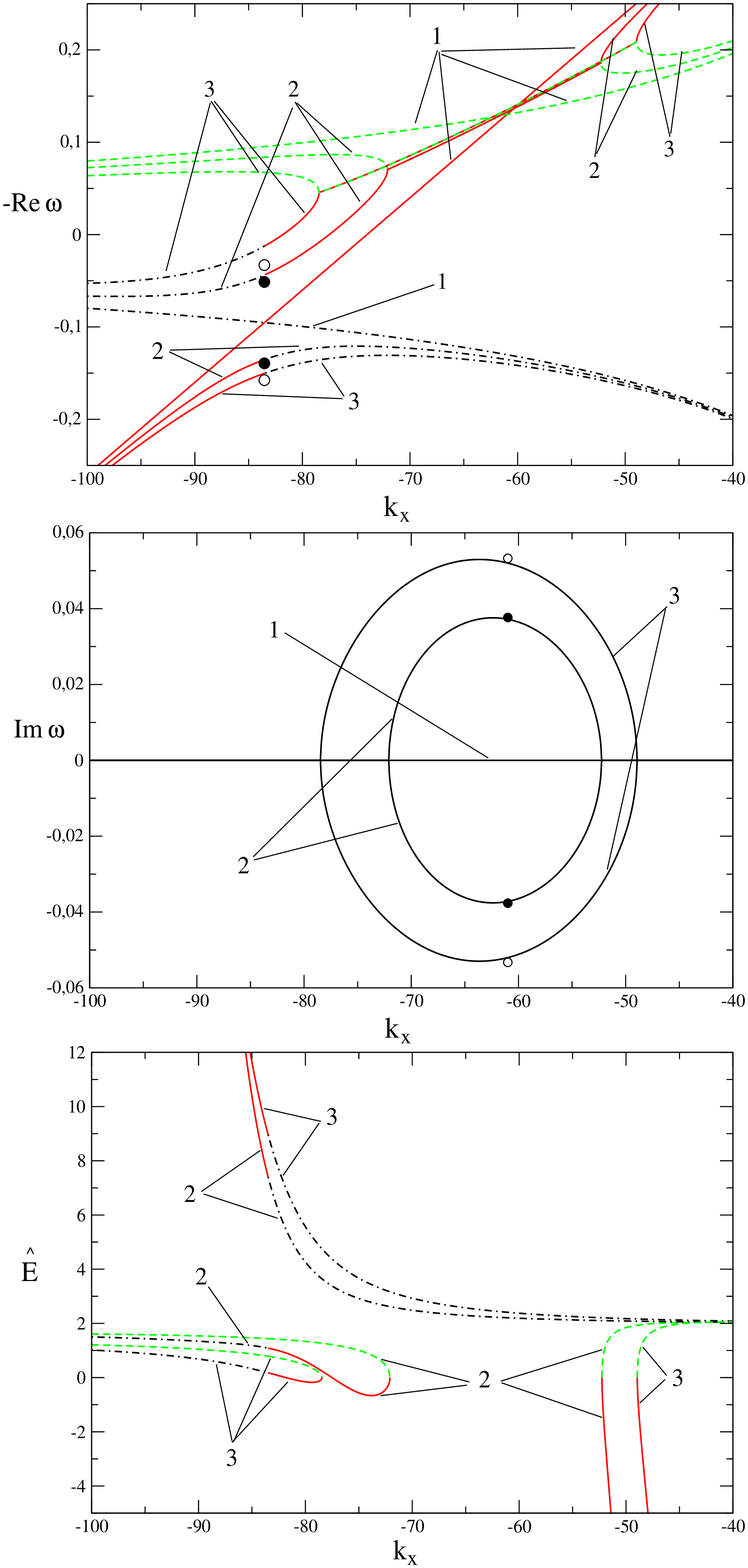}
\end{center}
\caption{
The mode coupling and the avoided crossing produced by the type-III and type-IV mode crossings, respectively.
Top panel: the curves show $\Re[\omega]$ taken with the opposite sign, where $\omega$ is the solution of the dispersion equation (\ref{disp}). The parameters are $q=3/2$, $\tau=0.1$, $g_x=0.1$, $g_z=0.925$ and $k_z=8$. 
The dust fraction is taken as $f=0$ (branches marked as '1'), 
$f=0.0005$ (branches marked as '2') and $f=0.001$ (branches marked as '3'). 
In the case $f=0$ the solid lines correspond to SDW, while the dashed and the dot-dashed lines correspond to IW propagating in the opposite directions.
In the case $f>0$ the similar lines correspond to modes akin to SDW and IW.
The filled and the hollow circles represent the analytical solution given by eq. (\ref{Delta_gen}) 
at the type-IV mode crossing according to eq. (\ref{gen_crossing}), respectively, for $f=0.0005$ and $f=0.001$.
Middle panel: the imaginary part of the corresponding solution.
The filled and the hollow circles represent the analytical solution given by eq. (\ref{Delta_gen}) at the type-III mode crossing
according to eq. (\ref{gen_crossing}), respectively, for $f=0.0005$ and $f=0.001$.
Bottom panel: the averaged energy density of modes with $\Im[\omega]=0$ as given by eq. (\ref{E_mode_varpi})  
for $|\varpi|^2=1$ and the non-zero $f$. 
The solid lines correspond to modes akin to SDW, while the dashed and the dot-dashed lines correspond to IW propagating in the
opposite directions.
}
\label{fig_5}
\end{figure}

In this Section the solution of eq. (\ref{disp}) is shown for the case when settling dominates the radial drift of the dust, 
see Fig. \ref{fig_4} for type-II crossing and Fig. \ref{fig_5} for type-III and type-IV crossings. 
The components of gravitational acceleration are chosen in such a way that the corresponding type-I crossing
occurs at the same point as in Fig. \ref{fig_1}. 
The mode coupling at the type-II crossing is similar to the case of dust settling, see Fig. \ref{fig_1},
with the difference that band of instability is slightly wider at the type-II crossing for the same value of the dust fraction.
In contrast, type-III crossing provides the band of instability in much greater range of $k_x$ with the growth rate by 
more than two times higher for the same value of the dust fraction. 
The bottom panel in Fig. \ref{fig_5} shows that in the vicinity of the mode coupling the energy of modes akin to IW and SDW
still have the opposite signs, however, as soon as $\Re[\omega]$ of the mode akin to SDW becomes positive, its energy also
acquires the positive values. Thus, the type-IV crossing of SDW which occurs with the other IW, provides the avoided crossing with
no SI. As a result, the loop of the energy profile forms to the left from the mode coupling in Fig. \ref{fig_5}. 
The analytical estimates made at type-III and type-IV crossings show less accordance with an accurate solution which may be 
caused by the proximity of the crossings to each other.
It can be checked that for greater value of the dust fraction, or alternatively, for higher $k_z$ than shown in Fig. \ref{fig_5} 
the mode coupling expands beyond the type-IV crossing, while at the same moment the ordinary coupling of two modes is replaced 
by the novel variant of coupling between three modes, i.e. two IW and one SDW, see Section \ref{sec_triple}.

\begin{figure}
\begin{center}
\includegraphics[width=8cm,angle=0]{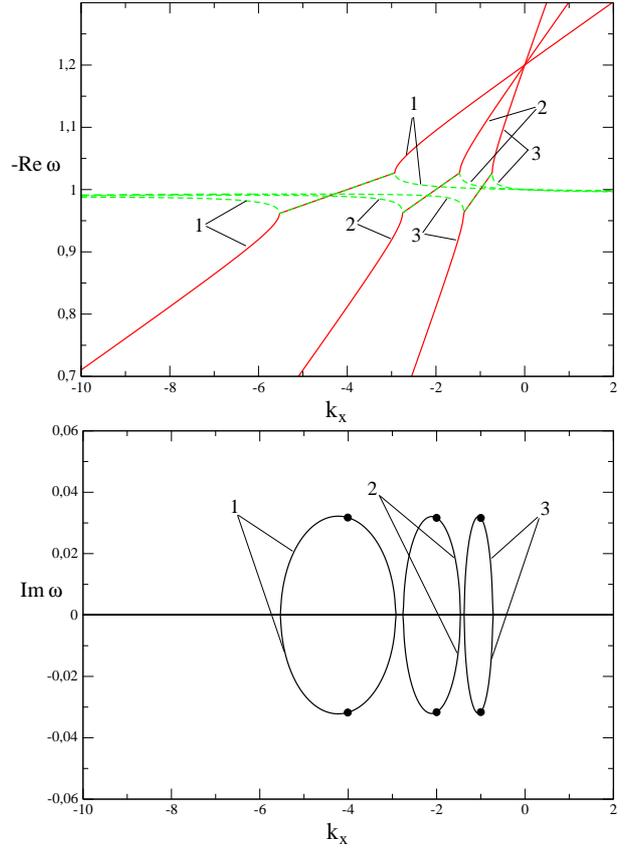}
\end{center}
\caption{
SI in the case of the dust radial drift dominating the dust settling.
Top panel: the curves show $\Re[\omega]$ taken with the opposite sign, where $\omega$ is the solution of the dispersion equation (\ref{disp}). The parameters are $q=3/2$, $\tau=0.1$, $g_z=0.1$, $f=0.01$ and $k_z=120$. 
The radial gravitational acceleration is taken as $g_x=0.5$ (branches marked as '1'), 
$g_x=1.0$ (branches marked as '2') and $g_x=2.0$ (branches marked as '3'). 
The solid and the dashed lines correspond to modes akin to SDW and IW, respectively.
Bottom panel: the imaginary part of the corresponding solution.
The filled circles represent the analytical solution given by eq. (\ref{Delta_gen}) at the mode crossing 
according to eq. (\ref{gen_crossing}).
}
\label{fig_6}
\end{figure}

In the opposite situation when the radial drift of the dust dominates its settling, i.e. $g_z \ll g_x$, type-III$^\prime$ crossing provides the mode coupling in the high-$k_z$ limit, see Fig. \ref{fig_6}. 
This case can be considered analytically employing the limits $g_x \gg g_z$ and $k_z \gg \tilde k_z$.
Then the growth rate given by eq. (\ref{Delta_gen}) is reduced to the following result
\begin{equation}
\label{Delta_fig_6}
\Delta \approx \pm {\rm i} \left ( \frac{f}{2} \right )^{1/2} (\kappa t_s g_z k_z)^{1/2}.
\end{equation}
Eq. (\ref{Delta_fig_6}) confirms that in this case the growth rate does not depend on the rate of radial drift, 
see Fig. \ref{fig_6}. Instead, the band of the instability shrinks over the radial wavenumber.
Eq. (\ref{Delta_fig_6}) can be compared with the growth rate in case of the dust settling, see eq. (\ref{Delta_2}),
$$
\Im[\omega] \sim \left ( \frac{k_z}{\tilde k_z} \right )^{1/2} \Im[\omega]_{\rm stl},
$$
which is formally even higher than the maximum value found in the absence of radial drift. 
However, as $g_z\to 0$, $\tilde k_z \to \infty$ and this branch of SI should be suppressed by viscosity in real disc.

\subsection{Triple coupling of modes}
\label{sec_triple}

\begin{figure}
\begin{center}
\includegraphics[width=8cm,angle=0]{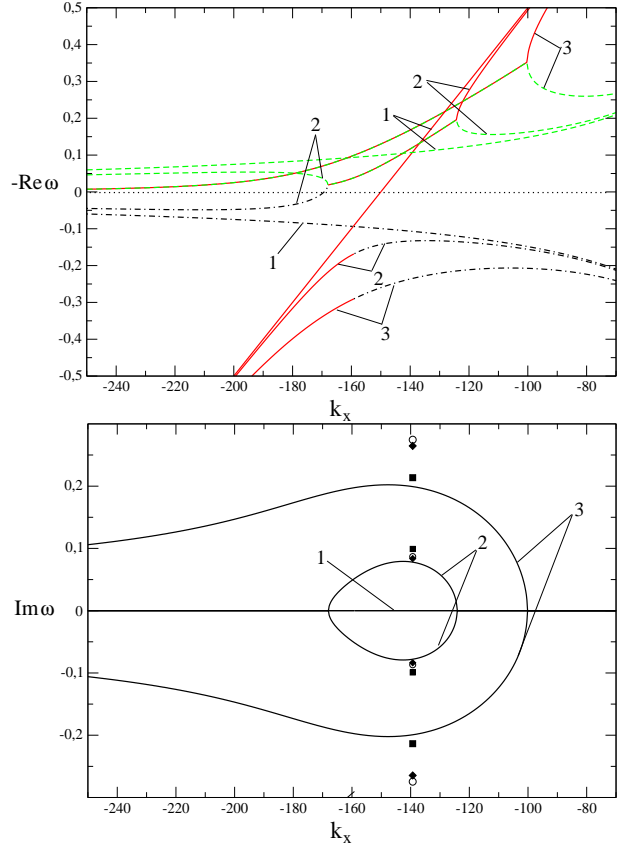}
\end{center}
\caption{
Illustration of the mode coupling, which produces SI with an unbounded growth rate.
Top panel: the curves show $\Re[\omega]$ taken with the opposite sign, where $\omega$ is the solution of the dispersion equation (\ref{disp}). The parameters are $q=3/2$, $\tau=0.1$, $g_x=0.1$, $g_z=1.0$ and $k_z=15$. 
The dust fraction is taken as $f=0$ (branches marked as '1'), 
$f=0.001$ (branches marked as '2') and $f=0.01$ (branches marked as '3'). 
In the case $f=0$ the solid lines correspond to SDW, while the dashed and the dot-dashed lines correspond to IW propagating in the opposite directions.
In the case $f>0$ the similar lines correspond to modes akin to SDW and IW.
Bottom panel: the imaginary part of the corresponding solution.
The filled diamonds represent the analytical solution given by eq. (\ref{Delta_gen}) at the mode crossing 
according to eq. (\ref{gen_crossing}). 
The hollow circles represent the approximate analytical solution given by eq. (\ref{1_2}).
The filled squares represent the approximate analytical solution given by the imaginary part of eq. (\ref{Delta_triple}).
All analytical estimations are made for $f=0.001$ and $f=0.01$. 
}
\label{fig_7}
\end{figure}

Here the mode coupling provided by the type-III crossing is considered in the high-wavenumber limit $k_z > \tilde k_z$ in the 
case when settling dominates radial drift, see Fig. \ref{fig_7}. As the dust fraction increases, the growth rate of SI behaves in 
a different way than it is expected from the ordinary mode coupling. 
In the high-$k$ limit the resonance condition (\ref{gen_crossing}) gives that $k_x \to - (g_z/g_x) k_z$ and eq. (\ref{Delta_gen}) yields
\begin{equation}
\label{1_2}
\Delta = \pm {\rm i} \left ( \frac{f}{2} \right )^{1/2} (\kappa t_s g_z k)^{1/2}. 
\end{equation}
As it is seen on the bottom panel in Fig. \ref{fig_7}, eq. (\ref{1_2}) along with the more general eq. (\ref{Delta_gen}) 
provides satisfactory approximation of an accurate solution of eq. (\ref{disp}) for smaller $f=0.001$ only. 
As the dust fraction becomes higher, those equations overestimate actual growth rate evaluated at the type-III crossing. 
The reason for such discrepancy can be found out deriving the approximate solution to eq. (\ref{disp}) in a slightly different way
than it is done in Section \ref{sec_mode_coupling}.
Indeed, eq. (\ref{disp}) can be written in another form:
\begin{equation}
\label{triple_disp}
(\omega - \omega_i) (\omega + \omega_i) (\omega - \omega_p) = \epsilon.
\end{equation}
Again, let the deviation from the type-III crossing caused by the coupling term be $\Delta = \omega - \omega_i = \omega - \omega_p$, where $\omega_c = \omega_i = \omega_p <0$.
As far as 
\begin{equation}
\label{triple_cond}
|\Delta| \ll |2\omega_i|,
\end{equation}
$\Delta$ is determined by equation
\begin{equation}
\label{Delta_1_2}
\Delta \approx \pm \left ( \frac{\epsilon}{2\omega_i} \right )^{1/2},
\end{equation}
which is equivalent to eq. (\ref{Delta}).

The condition (\ref{triple_cond}) implies that the correction to $\omega_c$ with the account of coupling is much less than 
the difference between the frequencies of oppositely propagating IW. However, as $|k_x|,k_z \to \infty$, 
this difference tends to constant value $\sim \kappa g_x/g_z$, while the coupling term $|\epsilon| \sim O(k) \to \infty$ 
as far as the dust fraction is constant.
In the marginal case
\begin{equation}
\label{triple_cond_marg}
|\Delta| \sim |2\omega_i|.
\end{equation}
Employing eq. (\ref{gen_crossing}) along with eq. (\ref{1_2}), one obtains from eq. (\ref{triple_cond_marg}) the upper value of $f$
for the ordinary mode coupling
\begin{equation}
\label{f_cr}
f_{cr} \approx \frac{8\kappa}{t_s k} \frac{g_x^2}{g_z g^2}.
\end{equation}
As $f>f_{cr}$, it does not occur any more.
In the opposite case $|\Delta| \gg |2\omega_i|$,
$\Delta$ is determined by equation
\begin{equation}
\label{Delta_1_3}
\Delta = \epsilon^{1/3},
\end{equation}
which yields the following result for the growth rate of the high-$k$ SI for $f>f_{cr}$:
\begin{equation}
\label{Delta_triple}
\Delta \approx \left \{ 1, -\frac{1}{2} \pm {\rm i} \frac{\sqrt{3}}{2} \right \} \left ( ft_s \kappa^2 k \frac{g_x g_z}{g} \right )^{1/3},
\end{equation}
which recovers an equation (5.16) of \citetalias{squire_2018}. 
Fig. \ref{fig_7} demonstrates that imaginary part of eq. (\ref{Delta_triple}) is in much better agreement with an accurate solution 
rather than eq. (\ref{1_2}). Note that for the case shown in fig. \ref{fig_7} $f_{cr} \approx 0.006$.

The transition from eq. (\ref{Delta_1_2}) to eq. (\ref{Delta_1_3}) implies than coupling of two modes is replaced by 
coupling of three modes, which are two IW and SDW. This new mechanism of instability, which can be referred to as
the triple mode coupling, is responsible for different dependence of RDI growth rate on the dust fraction, 
i.e. $\Im[\omega] \sim O(f^{1/3})$ instead of the more common $\Im[\omega] \sim O(f^{1/2})$. Eq. (\ref{f_cr}) shows that
the triple mode coupling is preferential for high wavenumbers in situations with the substantial settling of dust. Additionally, 
it occurs for particles of a greater size, as $f_{cr} \sim t_s^{-1}$. The weaker dependence of the growth rate on the small dust 
fraction can make the triple mode coupling the leading mechanism for dust clumping in certain regimes, which is advisable to check 
in the future work.

\subsection{Bonding of inertial waves}

\label{sec_bond}

It can be seen that in the case shown in fig. \ref{fig_7} the band of instability for $f=0.01$ extends unusually far towards 
$k_x\to -\infty$. It is shown below that this occurs due to the new instability of quasi-resonant nature.

\begin{figure}
\begin{center}
\includegraphics[width=8cm,angle=0]{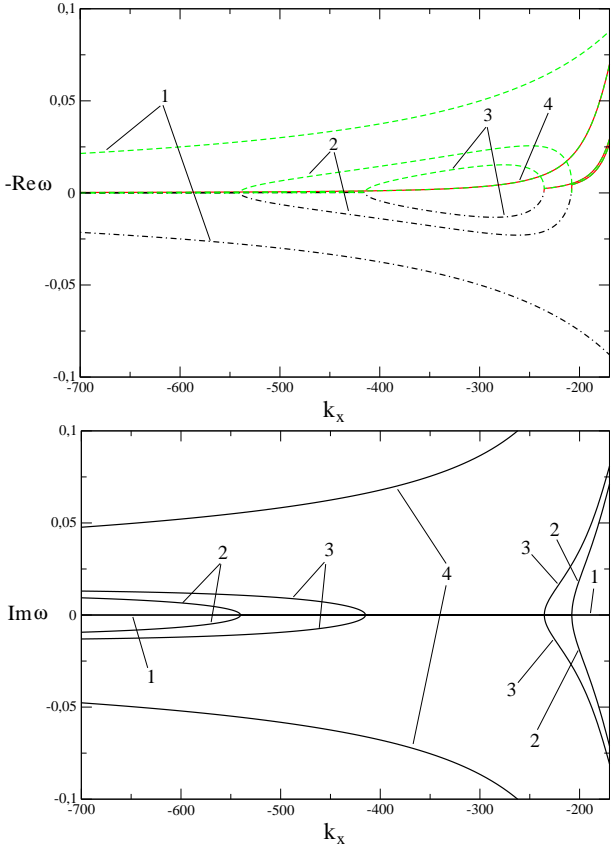}
\end{center}
\caption{
Bonding of low-frequency inertial waves propagating in the opposite directions.
Top panel: the curves show $\Re[\omega]$ taken with the opposite sign, where $\omega$ is the solution of the dispersion equation (\ref{disp}). The parameters are $q=3/2$, $\tau=0.1$, $g_x=0.1$, $g_z=1.0$ and $k_z=15$. 
The dust fraction is taken as $f=0$ (branches marked as '1'), 
$f=0.002$ (branches marked as '2'), $f=0.0023$ (branches marked as '3') and $f=0.01$ (branches marked as '4'). 
In the case $f=0$ the dashed and the dot-dashed lines correspond to IW propagating in the opposite directions.
In the case $f>0$ the similar lines correspond to modes akin to IW.
Bottom panel: the imaginary part of the corresponding solution.
}
\label{fig_8}
\end{figure}

The zone of lower negative $k_x$ is shown in Fig. \ref{fig_8} for the parameters used before in Fig. \ref{fig_7}.
As one can see, a band of new instability appears to the left from the band of SI produced by the type-III crossing.  
The top panel of fig. \ref{fig_8} shows that as this new instability sets on, the phase velocity of both modes akin to IW vanishes.
In that sense it resembles the mode coupling, however, the other features are different: \\
i) the mode crossing between two IW is located formally at infinity, $k_x\to -\infty$;\\
ii) both IW have positive energy at $k_x\to -\infty$; also, it can be checked that, as $f>0$, 
both modes akin to IW have positive energy outside of the band of new instability. \\
By these reasons, it is preferable to use a different term for this mechanism of instability. 
Hereafter it is referred to as 'bonding of IW'.

Thus, as $k_x\to -\infty$, two IW propagating in the opposite directions become the low-frequency modes. 
The difference between their phase velocities becomes small, which makes possible the bonding of IW caused by the coupling term, 
i.e. induced physically by the drag force acting on gas. 
Analytical estimation for the growth rate of bonding instability of IW can be obtained along the following lines.
Similar to the case with  the triple mode coupling, the starting point is eq. (\ref{triple_disp}).
Now, the substitution is $\omega = \omega_b + \Delta$, 
where $\omega_b=0$ is {\it not} the frequency of the mode crossing as in the case of the mode coupling, 
but rather the average of the opposite frequencies of two IW taken in the case $f=0$. 
As far as $|\Delta| \ll |\omega_p|$, this yields
\begin{equation}
\label{bond_disp}
\Delta^2 - \omega_i^2 = -\frac{\epsilon}{\omega_p}.
\end{equation}
Clearly, for the range of $k_x<0$ lower that the type-IV crossing both the coupling term $\epsilon$ and the frequency of SDW $\omega_p$
are positive, which implies that the bonding instability sets on as soon as
\begin{equation}
\label{bond_cond}
\omega_i^2 <  \left | \frac{\epsilon}{\omega_p} \right |.
\end{equation}
The equality in the instability condition (\ref{bond_cond}) yields the following approximate location of the bound of 
bonding instability along the radial wavenumbers:
\begin{equation}
\label{bond_bound}
\frac{k_x}{k_z} \approx -\frac{g_x}{f g_z} 
\end{equation}
provided that there is a sufficient dust radial drift:
\begin{equation}
\label{appr_bond_cond}
g_x \gg 2 f^{1/2} g_z.
\end{equation}
Note that $|k_x|\gg k_z$ in this limit.

As far as the absolute value of $k_x/k_z$ is much higher than the corresponding absolute value of (\ref{bond_bound}), the 
estimation of the growth rate is
\begin{equation}
\label{bond_growth_rate}
\Delta = \pm {\rm i} \kappa f^{1/2} \left ( \frac{g_z}{g_x} \frac{k_z}{|k_x|} \right )^{1/2}.
\end{equation}
As for RDI, the growth rate of bonding instability $\propto f^{1/2}$, which is a feature of resonant mechanism. 
However, here it is referred to as 'quasi-resonant' by the reasons mentioned above in this Section.

It was checked that eq. (\ref{bond_growth_rate}) gives a slight overestimation of an accurate value, nevertheless, 
the both values approach each other while $|k_x| \to \infty$. The frequency of corresponding gas-dust mode is imaginary, 
which means that bonding instability is provided by standing exponentially growing perturbations.

By virtue of eq. (\ref{bond_bound}) one can put the upper limit on the growth rate of the bonding instability as
\begin{equation}
\label{bond_restrict}
\Im[\omega] \lesssim \kappa f \frac{g_z}{g_x},
\end{equation}
which shows that it scales with the first power of the dust fraction, but still is comparable to SI growth rate 
in situation when dust settling dominates the radial drift of the dust. 
Eq. (\ref{bond_restrict}) implies that in geometrically thin disc the latter occurs as long as $h/r_0 \lesssim f^{1/2}$, which 
is commonly the case.
It is also important that, similar to SI, bonding instability does not depend on $t_s$.


\section{RDI beyond TVA}

\label{sec_beyond_TVA}

The analysis performed above shows that there is no RDI\footnote{as well as the bonding instability} 
in the absence of settling, $g_z\to 0$, within TVA.
The solutions of eq. (\ref{disp}) become symmetric with respect to change $k_x\to -k_x$, so that both type-I and type-IV crossings 
provide the identical avoided crossing, see eq. (\ref{Delta_real}). Hereafter it is assumed that $k_x>0$.
Thus, in order to recover RDI obtained by \citetalias{squire_2018} in the case of the dust radial drift with no settling, 
the next order terms over the small $\tau_*$ and $\lambda^{-1}$ must be retained.
Note that this concerns not only the terms $\sim UV$, but also the smallest terms $\sim V^2$ by the following reasoning.
Since the condition of the mode crossing is now
\begin{equation}
\label{res_cond_rad}
t_s g_x k_x = \frac{k_z}{k} \kappa,
\end{equation}
by the order of magnitude
$$
t_s g / l_{ev} \sim \Omega_0
$$
in the vicinity of resonance,
which implies that $\lambda^{-1} \sim \tau_*$, i.e. the terms $\sim V^2$ in eq. (\ref{eq_U}-\ref{eq_V}) are comparable 
to the inertial terms in eq. (\ref{eq_V}).

\subsection{Stationary solution}

Given the vertical gravity be negligible, $g_z=0$, the general set of equations (\ref{eq_U}-\ref{eq_rho_tot}) have the following stationary solution

\begin{equation}
\label{gen_stat_1}
{\bf U} = 0,
\end{equation}

\begin{equation}
\label{gen_stat_2}
\frac{\nabla (p+p_0)}{\rho} = -g_x {\bf e}_x,
\end{equation}

\begin{equation}
\label{gen_stat_3}
\rho_p = const,
\end{equation}

\begin{equation}
\label{gen_stat_4}
{\bf V} = -t_s g_x {\bf e}_x + t_s^2 g_x {\bf e}_y,
\end{equation}
where the terms $\sim \tau^3$ and smaller have been omitted.

\subsection{Reduced dispersion equation}

Here the strategy is to rederive the dispersion equation from the general set of equations 
for axisymmetric perturbations relaxing TVA, though adding the corresponding 
higher-order terms in $\tau$ in the coupling term only. For this reason, it is referred to 
as the reduced dispersion equation. The higher-order corrections either to $D_g$ or to $D_p$ are not considered as those
are not relevant to RDI, modifying the non-resonant dynamics of modes akin to IW or SDW and introducing (probably imaginary)
corrections to the frequencies $\omega_i$ and $\omega_p$, which are $\sim f$, rather than $\sim f^{1/2}$ as for the RDI 
growth rate.

The set of full equations (\ref{eq_U}-\ref{eq_rho_tot}) yields the following
equations for modes of linear gas-dust perturbations provided that the stationary solution is given 
by eqs. (\ref{gen_stat_1}-\ref{gen_stat_4})

\begin{equation}
\label{gen_mode_eq_1}
-{\rm i} \omega \hat u_x - 2\Omega_0 \hat u_y + F_x  = -{\rm i} k_x \hat W - f g_x \hat \delta,
\end{equation}

\begin{equation}
\label{gen_mode_eq_2}
-{\rm i} \omega \hat u_y + \frac{\kappa^2}{2\Omega_0} \hat u_x + F_y  = 0,
\end{equation}

\begin{equation}
\label{gen_mode_eq_3}
-{\rm i} \omega \hat u_z + F_z = -{\rm i} k_z \hat W,
\end{equation}

\begin{equation}
\label{gen_mode_eq_4}
-{\rm i} \omega \hat v_x - 2\Omega_0 \hat v_y - {\rm i} k_x t_s g_x \hat u_x + G_x = 
{\rm i} k_x \hat W + f g_x \hat \delta - \frac{\hat v_x}{t_s},
\end{equation}

\begin{equation}
\label{gen_mode_eq_5}
-{\rm i} \omega \hat v_y + \frac{\kappa^2}{2\Omega_0} \hat v_x - {\rm i} k_x t_s g_x \hat u_y + G_y = 
- f t_s g_x \frac{\kappa^2}{2\Omega_0} \hat \delta - \frac{\hat v_y}{t_s},
\end{equation}

\begin{equation}
\label{gen_mode_eq_6}
-{\rm i} \omega \hat v_z - {\rm i} k_x t_s g_x \hat u_z + G_z = {\rm i} k_z \hat W - \frac{\hat v_z}{t_s},
\end{equation}

\begin{equation}
\label{gen_mode_eq_7}
k_x \hat u_x + k_z \hat u_z = f ( k_x \hat v_x + k_z \hat v_z ) - ft_s g_x k_x \hat \delta,
\end{equation}
\begin{equation}
\label{gen_mode_eq_8}
-f \omega \hat \delta + k_x \hat u_x + k_z \hat u_z = 0,
\end{equation}
where it is assumed that $f\ll 1$, the terms $\sim f^2$ have been omitted and
\begin{equation}
\label{def_F}
{\bf F} \equiv f V_x  {\rm i} k_x ( \hat \delta \, {\bf V} + 2 {\bf \hat v})
\end{equation}
\begin{equation}
\label{def_G}
{\bf G} \equiv \frac{1}{(1+f)^2} V_x {\rm i} k_x ({\bf \hat v} - f \hat \delta {\bf V}). 
\end{equation}

The further derivation proceeds along the lines which lead to eqs. (\ref{mode_eq_1}-\ref{mode_eq_4}).
That is, the coupling term is a product of combinations of terms containing $\hat\delta$ in eqs. (\ref{gen_mode_eq_1}-\ref{gen_mode_eq_3}) and the term containing $\hat u_y$ in eq. (\ref{gen_mode_eq_8}), after the latter has been 
rearranged with the help of eqs. (\ref{gen_mode_eq_4}-\ref{gen_mode_eq_7}).

First, the components of perturbation of the relative velocity can be represented in the following way 
\begin{equation}
\label{gen_v_x}
\hat v_x \approx t_s\, [ 2\Omega_0 \frac{k_x^2}{k^2} \hat u_y - {\rm i} \omega_p \hat u_x ] (1 + {\rm i}t_s \omega - {\rm i}t_s \omega_p) - 2{\rm i} \, \Omega_0 t_s^2 \omega_p \hat u_y  + O(f \hat\delta),
\end{equation}

\begin{equation}
\label{gen_v_z}
\hat v_z \approx t_s\, [ 2\Omega_0 \frac{k_x k_z}{k^2} \hat u_y - {\rm i} \omega_p \hat u_z ] (1 + {\rm i}t_s \omega - {\rm i}t_s \omega_p)  + O(f \hat\delta),
\end{equation}
where it is assumed that 
\begin{equation}
\label{gen_om_p}
\omega_p = -t_s g_x k_x 
\end{equation}
and the higher order terms $\sim \tau^3$ have been omitted. 
Therefore, for purpose of this Section the term containing ${\bf v}$ in eq. (\ref{def_F}) can be omitted hereafter.
Eqs. (\ref{gen_v_x}-\ref{gen_v_z}) are substituted into eq. (\ref{gen_mode_eq_7}), so the second terms in square brackets 
of eqs. (\ref{gen_v_x}-\ref{gen_v_z}) yield together a single term $\sim O(f)$ which is also omitted in the new equation 
for $\hat \delta$.

At last, the set of eqs. (\ref{gen_mode_eq_1}-\ref{gen_mode_eq_8}) is rearranged as the following
\begin{equation}
\label{Gen_mode_eq_1}
-{\rm i} \omega \hat \phi - \omega k_x \hat u_z + 2\Omega_0 \hat \varpi = - {\rm i} f g_x k_z (1 - {\rm i} t_s \omega_p) \hat \delta,
\end{equation}
\begin{equation}
\label{Gen_mode_eq_2}
-{\rm i} \omega \hat \varpi - \frac{\kappa^2}{2\Omega_0} \hat \phi = 0,
\end{equation}
\begin{equation}
\label{Gen_mode_eq_3}
\omega k_x \hat \varpi - \frac{\kappa^2}{2\Omega_0} k_z^2 \hat u_z = - f k_z \omega \frac{\kappa^2}{2\Omega_0} \hat \delta,
\end{equation}
\begin{equation}
\label{Gen_mode_eq_4}
(\omega -\omega_p) k_z \hat \delta - 2 \tau (1 + {\rm i}t_s \omega - 2{\rm i } t_s \omega_p)\, {\rm i}k_x \hat \varpi = 0.
\end{equation}

Eqs. (\ref{Gen_mode_eq_1}-\ref{Gen_mode_eq_4}) lead to the reduced dispersion equation

\begin{equation}
\label{gen_disp}
D_g(\omega,{\bf k}) \cdot D_p(\omega,{\bf k}) = \tilde \epsilon (\omega,{\bf k}),
\end{equation}
where
\begin{equation}
\label{t_eps}
\tilde \epsilon \equiv -t_s f \kappa^2 k_x g_x \frac{k_z^2}{k^2} ( 1 + {\rm i} t_s \omega - 2{\rm i} t_s \omega_p )
\left ( 1 + \frac{{\rm i} k_x \omega^2}{g_x k_z^2} - {\rm i} t_s \omega_p \right )
\end{equation}
and it is assumed that $\omega_p$ is given by equation (\ref{gen_om_p}). 
Clearly, $\tilde \epsilon \to \epsilon$ in the leading order in $\tau$.

In turns out that in the vicinity of resonance (\ref{res_cond_rad}) the solution of equation (\ref{t_eps}) 
is in a good agreement with the solution of the full dispersion equation derived from equations (\ref{gen_mode_eq_1}-\ref{gen_mode_eq_8}), see Fig. 1 of \citetalias{squire_2018}.

The approximate solution of eq. (\ref{disp}) at the mode crossing of IW and SDW given by eq. (\ref{Delta}) 
with the replacement $\epsilon \to \tilde \epsilon|_{\omega_c}$ applied to equation (\ref{gen_disp}) yields
\begin{equation}
\label{gen_approx_sol}
\Delta \approx \pm \kappa \left ( \frac{f}{2} \right )^{1/2} \frac{k_z}{k} 
\left [ 1 + {\rm i} \frac{t_s \kappa}{2} \frac{k_z}{k} \left ( 2 + \frac{k_x^2}{k_z^2} \right ) \right ],
\end{equation}
where the resonance condition (\ref{res_cond_rad}) is used. 
Eq. (\ref{gen_approx_sol}) recovers the result of \citetalias{squire_2018}, see their equation (5.10).

Consideration of RDI of radially streaming dust reveals that the underlying mechanism goes beyond the concept of 
the mode coupling. In this situation, RDI is caused by a quasi-resonant process manifested
in the growth/damping of the uncoupled modes both having positive energy and slightly different phase velocities at the 
mode crossing. Moreover, as the dust fraction increases, the growth rate increases along with 
the degree of the decoupling between IW and SDW, see eq. (\ref{gen_approx_sol}).
At the same time, such a mechanism is generally weaker than the mode coupling as it emerges due to inertia of solids 
represented by the next-order terms in $\tau$ entering an augmented coupling term (\ref{t_eps}).

\section{Conclusions}

This study is focused on the nature of local instability of gas-dust mixture with dust streaming in protoplanetary disc. 
Previously, \citetalias{squire_2018} revealed that in the leading order in small $f$ axisymmetric perturbations
grow due to RDI, which provides the growth rate $\propto f^{1/2}$.
Here RDI is reconsidered in terms of perturbations of the center-of-mass velocity and the relative velocity of 
gas-dust mixture as well as perturbation of dust density. This allows for the analysis of RDI within TVA.
The dynamics of modes of perturbations is considered in the general case of dust settling combined with dust radial drift 
see eqs. (\ref{mode_eq_1}-\ref{mode_eq_4}). 
The usefull framework of the Lagrangian perturbation theory for neutral modes of perturbations is applied to two-fluid model 
in order to show the existence of negative energy waves in such a flow from fundamental symmetry of the system with respect to 
translations in time, see eq. (\ref{E_mode_varpi}). 
Generally, the negative energy wave is akin to SDW, while it becomes identical to SDW in the absence of dust back reaction on gas, 
which is reproduced by the formal limit $f\to 0$. The energy of SDW is given by eq. (\ref{E_1_mode}).
In the particular case of the dust settling without radial drift, there is a straightforward generalisation 
of variational principle onto arbitrary linear gas-dust perturbations, which makes it possible to show that 
the energy of growing/damping modes vanishes, see the Appendix \ref{App_2}.

The dispersion equation (\ref{disp}) for gas-dust perturbations with the account for dust back reaction on gas is reduced to form typical for systems with the mode coupling. In the absence of the coupling term the dispersion equation splits into two independent
equations describing one SDW and two oppositely propagating IW, respectively. As the dispersion curve of SDW crosses the dispersion
curve of IW, the modes fall in resonance, and the corresponding point of phase space is referred to as the mode crossing. 
There can be up to four mode crossings for the given free parameters of the model, see Fig. \ref{fig_2}. 
The formal classification of mode crossings is suggested in table \ref{types}. Their existence in phase space is outlined 
in table \ref{existence}. Type-I crossing occurs for any combination of settling and radial drift, though in the limited 
range of wavenumbers. Type-II crossing occurs in the limited range of wavenumbers, but for the substantial settling only. 
Type-III crossing occurs for the substantial settling only, but at sufficiently high wavenumbers. 
Type-III$^\prime$ crossing replaces type-III crossing as radial drift of the dust dominates its settling. 
At last, type-IV crossing occurs for 
all wavenumbers each time there is a radial drift of the dust. In the marginal case of dust streaming only vertically
there are type-I,II crossings identical to each other, whereas, in the marginal case of dust streaming only radially
there are type-I,IV crossings identical to each other.

The definite criterion for the existence of SI, i.e. RDI in the presence of the dust settling, 
at the given mode crossing is formulated in Section \ref{sec_mode_coupling}:  
the energy of SDW evaluated at the mode crossing must be negative. In the opposite case of the positive energy of SDW, 
the avoided crossing occurs, which gives no instability within TVA. 
The avoided crossing takes place for type-IV crossing and also for type-I crossing in the case of a sufficient 
radial drift of solids, see the conditions (\ref{avoided_k_x}-\ref{avoided_k_z}). 
In other situations the mode crossing provides SI.
SI occurs due to the energy transfer from (negative energy) SDW to (positive energy) IW in the vicinity of the corresponding mode crossing. The coupled modes are represented by the complex conjugate pair of frequencies obtained from 
the accurate solution of the dispersion equation, see Figs. (\ref{fig_1},\ref{fig_3}-\ref{fig_6}).
The band of instability expands in the space of wavevectors as the coupling term in the dispersion equation, 
which is proportional to the dust mass fraction, becomes larger. 
SI attains the maximum growth rate approximately at the mode crossing. It is estimated analytically employing the usual
mode coupling approach, see eq. (\ref{Delta}). Eq. (\ref{Delta}) also indicates that within TVA 
the correction to the mode frequency due to the dust back reaction on gas is generally independent on $t_s$. The latter
applies to both the mode coupling and the avoided crossing.

The branch of SI having an unbounded asymptotics of growth rate at high wavenumbers is provided by mode coupling 
at type-III($^\prime$) crossing. The corresponding analytical estimate of maximum growth rate is given by eq. (\ref{1_2}).
Note that it is determined solely by the rate of vertical settling $\sim t_s g_z$ despite the presence of radial drift
necessary for the very existence of the considered mode coupling. Also, this branch of SI takes place for wavenumbers of different 
signs, i.e. either for $k_z>0$ and $k_x<0$, or for $k_z<0$ and $k_x>0$. Further, as far as $k$ increases for constant dust fraction,
the distance between type-III and type-IV crossings in phase space tends to a constant value, 
whereas the absolute value of the coupling term grows up. Eventually, the ordinary coupling between SDW and one of IW is replaced
by the triple coupling between SDW and the both of IW, see Section \ref{sec_triple} for details. As a result, the growth rate acquires
a different dependence on the dust fraction, $\propto f^{1/2\to 1/3}$. The triple mode coupling yields eq. (\ref{Delta_triple})
for estimation of the maximum growth rate, which recovers the result of \citetalias{squire_2018}. Note that unlike 
eq. (\ref{1_2}) it explicitly depends on both components of gravitational acceleration.

For the absolute value of radial wavenumber even higher than that of type-III,IV crossings, an additional branch of instability 
emerges, see Section \ref{sec_bond} for details. For the sufficiently high dust fraction it is attached to the band of
instability provided by the mode coupling at type-III crossing, see the curve marked `3' on the bottom panel of Fig. \ref{fig_7}.
As is shown in Fig. \ref{fig_8}, the new instability appears in the region where two low-frequency oppositely 
propagating IW coalesce with each other, while their phase velocity vanishes. 
As in the case of the mode coupling, the dust back reaction on gas, which introduces the coupling term of the dispersion equation, 
is responsible for this new process. However, the underlying mechanism of growth is different, since the coalesced modes have positive energy and there is no resonance between the modes in the limit $f\to 0$. Nevertheless, the low-frequency IW approach each other
in phase space as $|k_x|\to \infty$, while the growth rate of new instability $\propto f^{1/2}$, see eq. (\ref{bond_growth_rate}).
By these reasons, in this study it is referred to as quasi-resonant instability of bonding IW.
The upper limit of the bonding instability estimated by eq. (\ref{bond_restrict}) confirms that this instability is 
important along with SI in the situation when the settling of the dust dominates its radial drift. 

The short wavelengths typical for both the triple mode coupling and the bonding instability are susceptible to dissipation 
effects in a disc. The influence of effective viscosity of gas on these branches of instability should be considered in the
future work in order to understand their relevance for structure formation in real protoplanetary discs.  
Besides, moderate dissipation may not necessarily lead to damping of high-wavenumber SI. 
In contrast, it may cause an additional non-resonant growth of negative energy mode akin to SDW.
This issue will be addressed in the subsequent studies.

In the limit of the dust streaming only radially, $g_x \gg g_z$, or equivalently $g_z\to 0$, SI ceases to operate. 
The growth rate due to the mode coupling produced by type-I crossing vanishes according to eq. (\ref{avoided_Delta}), while 
the mode coupling produced by type-III($^\prime$) crossing shifts towards $k\to \infty$ to be suppressed by 
any small amount of viscosity, see eqs. (\ref{Delta_fig_6}), (\ref{1_2}) and (\ref{Delta_triple}).
Strictly for $g_z=0$, there exist type-I and type-IV crossings identical to each other, which produce the avoided crossings
with the account for dust back reaction on gas. Thus, there is no RDI within TVA when dust is subject only to radial drift.
RDI is recovered by retaining the terms of the orders of $\tau_*$, $\sqrt{\lambda^{-1}}$ and $\lambda^{-1}$ from the 
full equations (\ref{eq_U}-\ref{eq_V}), which contribute to the coupling term of the dispersion equation, 
see eqs. (\ref{gen_disp}-\ref{t_eps}). It turns out that in the vicinity of resonance between SDW and IW all those terms 
yield the imaginary correction $\sim O(\tau)$ to the coupling term. Therefore, the two separate branches of neutral modes 
representing the avoided crossing acquire small growing/damping. In this way, it becomes clear that the streaming instability
of \citet{youdin-goodman-2005} is provided by resonant mechanism other than the mode coupling responsible for SI.
This new mechanism remains obscured. Though, it should be related to the inertia of solids, because the latter is exactly
what is neglected in TVA. Also, its relationship with the inertia of solids naturally explains the increase of the growth rate 
with the size of the particles.

The dispersion equation obtained beyond TVA, see eq. (\ref{gen_disp}), is a reduced one, since the corrections $\sim O(f)$ to 
the dispersion relations of IW and SDW given, respectively, by $D_g$ and $D_p$, have been omitted there. 
However, as shows the analytical study of \citet{latter-2011}, the modes can exhibit an intrinsic growth due 
to non-resonant mechanisms. This is especially probable for negative energy SDW, what remains to be checked further.
The non-resonant growth may be important as $f \to 1$.


The relevance of the streaming instability to planetesimal formation has being extensively studied 
via local simulations of the non-linear dynamics of perturbations in dust-laden disc midplane, see the recent results by \citet{johansen-2017}. 
Although the numerical models usually incorporate the dust settling, SI has not been found so far, see the discussion 
by \citet{squire_2018}. Besides, SI might be less sensitive to external turbulence inherent in astrophysical discs, see
the recent numerical study of the dust settling through turbulent gas by \citet{lin-2019}.
At last, it is highly likely that the other RDI revealed by \citet{squire_2018} are caused by coupling of SDW 
with waves of the other origin, such as sound waves and internal gravity waves ubiquitous in protoplanetary discs.
These issues are relegated to future work.


\section*{Acknowledgments}

The author acknowledges the support from the Program of development of M.V. Lomonosov Moscow State University (Leading Scientific School 'Physics of stars, relativistic objects and galaxies').

\bibliography{bibliography}



\appendix

\section{Equations for perturbations in TVA}
\label{App_1}

The perturbed state of gas-dust mixture can be described by 
the small Eulerian perturbations of the centre-of-mass velocity, ${\bf u}$, the relative velocity, ${\bf v}$, 
the gas pressure, $p^\prime$, and the density of dust, $\rho_p^\prime$.  
Those quantities obey the corresponding linear equations

\begin{equation}
\label{pert_U}
\begin{aligned}
(\partial_t - q\Omega_0 x \partial_y)\, {\bf u} - 2\Omega_0 u_y {\bf e}_x + (2-q)\Omega_0 u_x {\bf e}_y + \\ 
({\bf u}\nabla){\bf U} + ({\bf U}\nabla) {\bf u} =  
-\frac{\nabla p^\prime}{\rho} + \frac{\nabla (p+p_0)}{\rho} \frac{\rho_p^\prime}{\rho},
\end{aligned}
\end{equation}

\begin{equation}
\label{pert_V}
\frac{\nabla p^\prime}{\rho} - \frac{\nabla(p+p_0)}{\rho} \frac{\rho_p^\prime}{\rho} = \frac{{\bf v}}{t_s},
\end{equation}

\begin{equation}
\label{pert_rho}
\nabla \cdot \left ( {\bf u} - \frac{\rho_p}{\rho} {\bf v} - 
\frac{\rho_g}{\rho} \frac{\rho_p^\prime}{\rho} {\bf V} \right ) = 0,
\end{equation}

\begin{equation}
\label{pert_rho_tot}
(\partial_t - q\Omega_0 x \partial_y)\, \rho_p^\prime + \nabla (\rho_p^\prime {\bf U} + \rho\, {\bf u})  = 0.
\end{equation}

The set of eqs. (\ref{pert_U}-\ref{pert_rho_tot}) describes the evolution of gas-dust perturbations for any stationary 
solution of eqs. (\ref{eq_U_2}), (\ref{eq_TVA}), (\ref{eq_rho_g}) and (\ref{eq_rho_tot}) on the length-scales much shorter than the scaleheight of a thin protoplanetary disc.
Eqs. (\ref{pert_U}-\ref{pert_rho_tot}) become much more simple for the particular background model considered 
in Section \ref{gen_stat_sol}.

Once the background model is specified by eqs. (\ref{bg_U}-\ref{bg_sigma}),
one arrives at the following equations for perturbations
\begin{equation}
\label{u}
\begin{aligned}
(\partial_t - q\Omega_0 x \partial_y)\, {\bf u} - 2\Omega_0 u_y {\bf e}_x + 
(2-q)\Omega_0 u_x {\bf e}_y = \\ -\nabla W +\frac{f}{1+f} \delta {\bf g},
\end{aligned}
\end{equation}
\begin{equation}
\label{delta}
(\partial_t - q\Omega_0 x \partial_y)\, {\delta} = -t_s \nabla^2 W - \frac{1-f}{1+f} t_s ({\bf g} \nabla) \delta,
\end{equation}

\begin{equation}
\label{div_u}
\nabla \cdot {\bf u} = \frac{t_s f}{1+f} \nabla^2 W + t_s f \frac{1-f}{(1+f)^2} ({\bf g} \nabla) \delta,
\end{equation}
where $W \equiv p^\prime/\rho$ and the relative perturbation of the dust density $\delta \equiv \rho_p^\prime/\rho_p$.
Note that eqs. (\ref{u}-\ref{div_u}) are valid for $f$ not necessarily of the small value.

\subsection{Equations for axisymmetric perturbations in the leading order in small dust fraction}

Let additionally the dust fraction be small $f\ll 1$ and perturbations have axial symmetry.
Eqs. (\ref{u}-\ref{div_u}) yield
\begin{equation}
\label{u_x}
\partial_t u_x - 2\Omega_0 u_y = - \partial_x W - f g_x \delta,
\end{equation}
\begin{equation}
\label{u_y}
\partial_t u_y + (2-q)\Omega_0 u_x = 0,
\end{equation}
\begin{equation}
\label{u_z}
\partial_t u_z = - \partial_z W - f g_z \delta,
\end{equation}
\begin{equation}
\label{delta_2}
\partial_t \delta = - t_s (\partial^2_{xx} + \partial^2_{zz}) W + t_s ( g_z \partial_z \delta + g_x \partial_x \delta ),
\end{equation}
\begin{equation}
\label{div_u_2}
\partial_x u_x + \partial_z u_z = 0.
\end{equation}

Hence, in this the most simple case the small gas-dust perturbations of dust streaming through the static gas environment 
both vertically and radially are described by eqs. (\ref{u_x}-\ref{u_z}, \ref{div_u_2}) for perturbation 
of the centre-of-mass velocity, ${\bf u}$. These equations are identical to equations for single-fluid dynamics 
of vortical perturbations in a rotating plane shear flow, see e.g. equations (6-9) of \citet{umurhan-regev-2004} with $\partial_y=0$, 
except for the terms in the RHS of eqs. (\ref{u_x}) and (\ref{u_z}) proportional to perturbation of the dust density. 
At the same time, perturbation of the dust density is the only quantity that additionally describes the dynamics of dust 
by means of eq. (\ref{delta_2}). The perturbed flow of dust is affected by gas through the term $\propto \nabla^2 W$ in the RHS of eq. (\ref{delta_2}). 
In the limit $f\to 0$ the set of eqs. (\ref{u_x}-\ref{div_u_2}) splits into eqs. (\ref{u_x}-\ref{u_z}, \ref{div_u_2}), which 
determine the dynamics of solely the gas perturbations with ${\bf u}$ becoming identical to the velocity of gas,
and eq. (\ref{delta_2}), which separately determines the dynamics of dust perturbations. In the latter situation, solids 
move passively under the action of external gravity and aerodynamic drag. If additionally $t_s\to 0$, 
both the background relative velocity and the perturbation of relative velocity vanish, which results in the conservation of the
density of dust frozen in the divergence-free motion of gas.

The main results of this paper come out from eqs. (\ref{u_x}-\ref{div_u_2}).

\subsection{Equations for perturbations in the appropriate variables}

It is appropriate to reformulate eqs. (\ref{u_x}-\ref{div_u_2}) in terms of the variables
\begin{equation}
\label{chi}
\chi \equiv \{\varpi,\,\phi,\,u_z,\,\delta\}.
\end{equation}

It is assumed that components of $\chi$ have non-zero derivatives over $t,x,z$, which are denoted below as $\partial_k\chi_i$.
The usual Einstein's rule of summation over the repeated upper and lower indices is assumed as well. 

First, taking the curl of eqs. (\ref{u_x}-\ref{u_z}) yields the following
\begin{equation}
\label{sys_1}
\partial^2_{tz} u_x - \partial^2_{tx} u_z = 2\Omega_0 \partial_z u_y + f ( g_z \partial_x \delta - g_x \partial_z \delta ),  
\end{equation}
\begin{equation}
\label{sys_2}
-\partial^2_{tz} u_y = (2-q)\Omega_0 \partial_z u_x,
\end{equation}
\begin{equation}
\label{sys_3}
\partial^2_{tx} u_y =  (2-q)\Omega_0 \partial_z u_z.
\end{equation}

Note that eqs. (\ref{sys_1}-\ref{sys_3}) are identical to equations describing 
IW in rigidly rotating fluid provided that $q=f=0$, see e.g. \citet{landau-lifshitz-1987}, paragraph 14. 
The LHS of eqs. (\ref{sys_1}-\ref{sys_3}) are the time derivatives 
of components of vorticity perturbation\footnote{In this work vorticity is a curl of the center-of-mass velocity of gas-dust mixture rather than a curl of the velocity of gas}.

Next, the divergence of eqs. (\ref{u_x}-\ref{u_z}) yields
\begin{equation}
\label{nabla_W}
(\partial^2_{xx} + \partial^2_{zz}) W =  2\Omega_0 \partial_x u_y - f ( g_x \partial_x \delta + g_z \partial_z \delta ).
\end{equation}

Taking additional derivatives over $z$ from equations (\ref{delta_2}), (\ref{nabla_W}), (\ref{sys_3}) 
and employing eq. (\ref{div_u_2}) one comes to the final set of equations for the new variables

\begin{equation}
\label{Sys_1_A}
\partial_t \phi = \partial^2_{tx} u_z  - 2\Omega_0 \varpi + f (g_z \partial_x \delta - g_x \partial_z \delta),  
\end{equation}
\begin{equation}
\label{Sys_2_A}
\partial_t \varpi = \frac{\kappa^2}{2\Omega_0} \phi,
\end{equation}
\begin{equation}
\label{Sys_3_A}
\partial^2_{tx}\varpi = - \frac{\kappa^2}{2\Omega_0} \partial^2_{zz} u_z,
\end{equation}
\begin{equation}
\label{Sys_4_A}
\partial^2_{tz}\delta = t_s (g_z \partial^2_{zz}\delta + g_x \partial^2_{xz} \delta  ) + 2\tau \partial_x \varpi,
\end{equation}
where the term $\propto ft_s$ has been neglected in eq. (\ref{Sys_4_A}).

\section{Variational principle for dynamics of general gas-dust perturbations in the case of the dust vertical settling}
\label{App_2}

The set of eqs. (\ref{Sys_1_A}-\ref{Sys_4_A}) for $g_x=0$ can be derived from the requirement that action
\begin{equation}
\label{action}
S = \int {\cal L}(\chi_i, \partial_k \chi_i) d^2{\bf x}\, dt
\end{equation}
with the Lagrangian density
\begin{equation}
\label{L_full}
{\cal L} = {\cal L}_0 + f {\cal L}_1
\end{equation}
be stationary with respect to arbitrary variations of $\chi_i$ in the small shearing box. 
Similar to the variational principle for modes, see eq. (\ref{action_modes}),
the subscript `0' denotes the contribution to the full Lagrangian responsible for the dynamics of gas-dust perturbations 
with no account for the drag force acting on gas, while the subscript `1' denotes additional contribution to ${\cal L}$ responsible solely for the dynamics of dust.
If so, eqs. (\ref{Sys_1_A}-\ref{Sys_4_A}) are identical to the Euler-Lagrange equations
$$
\frac{\delta {\cal L}}{\delta \chi_i} - \partial_k \left ( \frac{\delta {\cal L}}{\delta (\partial_k \chi_i)} \right ) = 0, 
$$
where explicitly
\begin{equation}
\label{L_0}
{\cal L}_0 = - \varpi \partial_t\phi - \partial_t u_z \partial_x\varpi - \Omega_0\varpi^2 - \frac{\kappa^2}{2\Omega_0}\frac{\phi^2}{2} -
\frac{\kappa^2}{2\Omega_0} \frac{(\partial_z u_z)^2}{2},
\end{equation}
\begin{equation}
\label{L_1}
{\cal L}_1 = g_z \, \left\{ \varpi \partial_x\delta -  
\frac{1}{2\Omega_0} \, \left [ \frac{\partial_t\delta \partial_z\delta}{2t_s} - g_z \frac{(\partial_z\delta)^2}{2} \right ]\, \right\}.
\end{equation}
The first term in ${\cal L}_1$ is the generalized cross-term, cf. eq. (\ref{L_1_modes}).
Like the Lagrangian (\ref{L_full_modes}), $\cal{L}$ is known up to an arbitrary constant factor chosen
so that the energy of IW propagating in rigidly rotating fluid be positive.

\subsection{Energy of general gas-dust perturbations}

The energy of general perturbations associated with the invariance of ${\cal L}$ with respect to translations in time is
$E \equiv \int {\cal E} d^2{\bf x}$, where
$$
{\cal E} = -{\cal L} + \frac{\delta {\cal L}}{\delta (\partial_t \chi_i)} \partial_t \chi_i
$$
is the energy density. Similar to the energy density of mode (\ref{E_full_modes}), it consists of the main part recovering 
the energy density of general gas perturbations in rotating flow in the absence of dust, 
and additional contribution related to general perturbations of dust density
\begin{equation}
\label{E_full}
{\cal E} = {\cal E}_0 + f {\cal E}_1.
\end{equation}
Explicitly,
\begin{equation}
\label{E_0}
{\cal E}_0 = \Omega_0\varpi^2 + \frac{\kappa^2}{2\Omega_0} \frac{\phi^2}{2} + \frac{\kappa^2}{2\Omega_0} \frac{(\partial_z u_z)^2}{2},
\end{equation}
\begin{equation}
\label{E_1}
{\cal E}_1 = - g_z\, \left \{ \varpi \partial_x\delta + \frac{g_z}{2\Omega_0} \frac{(\partial_z\delta)^2}{2} \right \}.
\end{equation}

It is straightforward to check that eqs. (\ref{Sys_1_A}-\ref{Sys_4_A}) guarantee that $\partial_t {\cal E}$ can be represented as 
the divergence of vector, which is identified with the energy flux vanishing along with perturbations $\chi_i$
as one goes to infinity.
${\cal E}_0$ is of course a positive definite quantity in centrifugally stable flows, while 
${\cal E}_1$ can have any sign depending on the contribution
of the cross term. However, the net contribution of the cross term to $E$ can be small or even vanish for certain non-trivial perturbations, i.e. for modes, see eq. (\ref{E_full_modes}). 
Then, owing to perturbations of dust density, ${\cal E}_1$ gives negative amount to the energy for the case $g_x=0$ considered here.

\subsection{Energy of growing mode}

In order to obtain expression for the energy density of mode, which is not necessarily neutral, 
$\Re[\chi_i]$ is substituted into eq. (\ref{E_full}) 
and subsequently averaged over the wavelength of mode.
Eqs. (\ref{mode_eq_1}-\ref{mode_eq_4}) provide the following result
\begin{equation}
\label{E_mode}
\begin{aligned}
& \hat {\cal E} = |\hat \varpi|^2 {\rm e}^{2\Im [\omega] t} \left \{ 1 + \frac{(\Re[\omega])^2 + (\Im[\omega])^2}{\kappa^2} \frac{k^2}{k_z^2} + \right . \\ 
&\left . 
\frac{f g_z k_x^2}{(\Re[\omega] + t_s g_z k_z)^2 + (\Im[\omega])^2} \left[ \frac{2t_s}{k_z} (\Re[\omega] +t_s g_z k_z) - 
g_z t_s^2 \right]\,  \right \}, 
\end{aligned}
\end{equation}
where $k^2 \equiv k_x^2 + k_z^2$.

In this way, one obtains the averaged energy density of the mode of gas-dust mixture perturbations for 
a particular wavevector ${\bf k}$, provided that its (complex) frequency is known, i.e. there is a solution of a dispersion equation.

\subsubsection{Approximate solution at the mode crossing and the vanish of the mode energy}
\label{App2_2}

Eq. (\ref{disp}) is cubic with respect to $\omega$. Seeking the Cardano solution of eq. (\ref{disp}) for $g_x=0$ at the 
mode crossing (\ref{crossing}) one finds that the discriminant 
$$
Q= \frac{8}{27} f \kappa^6 \frac{k_x^2 k_z^4}{k^6} \left ( 1 + \frac{27}{32} f \frac{k_x^2}{k_z^2} \right )
$$ 
vanishes for $f=0$, and eq. (\ref{disp}) acquires a double root. Then, any small non-zero $f>0$ makes $Q$ be a positive
value, which implies that a double root of eq. (\ref{disp}) acquires the adjoint couple of imaginary parts, 
i.e. the instability sets on in some band around the mode crossing.

In the case of small non-zero $f>0$ the Taylor expansion of the Cardano solution of eq. (\ref{disp}) by the orders of $f^{1/2}$ 
at the mode crossing yields the following approximate real and imaginary parts of the frequency
\begin{equation}
\label{approx_re}
\frac{\Re[\omega]}{\omega_c} = 1 + \frac{f}{8} \frac{k_x^2}{k_z^2},
\end{equation}

\begin{equation}
\label{approx_im}
\frac{\Im[\omega]}{\omega_c} = \pm \left ( \frac{f}{2} \right )^{1/2} \frac{k_x}{k_z} \left ( 1 - \frac{5f}{64} \frac{k_x^2}{k_z^2} \right ).
\end{equation}
Note that eq. (\ref{approx_im}) recovers eq. (\ref{Delta_2}) in the leading order in $f^{1/2}$.
It is straightforward to check that expressions (\ref{approx_re}) and (\ref{approx_im}) make $\hat {\cal E}$ as given by 
eq. (\ref{E_mode}) vanish for any small $f>0$.

\end{document}